\documentclass[sn-mathphys-num]{sn-jnl}
\usepackage{fix-cm}

\usepackage{graphicx}%
\usepackage{multirow}%
\usepackage{amsmath,amssymb,amsfonts}%
\usepackage{amsthm}%
\usepackage{mathrsfs}%
\usepackage[title]{appendix}%
\usepackage{xcolor}%
\usepackage{textcomp}%
\usepackage{manyfoot}%
\usepackage{booktabs}%
\usepackage{algorithm}%
\usepackage{algorithmicx}%
\usepackage{algpseudocode}%
\usepackage{listings}%
\usepackage{fancyhdr}
\usepackage{natbib}
\usepackage{bm}

\newtheorem{theorem}{Theorem}
\newtheorem{corollary}[theorem]{Corollary}%

\renewcommand{\footnoterule}{%
	
	\hrule width 0.3\textwidth height 0.4pt
	\kern 6pt  
}

\newtheorem{remark}{Remark}%
\newtheorem{lemma}{Lemma}%

\newtheorem{definition}{Definition}%
\usepackage{hyperref}
\begin{document}
	
	\title[Article Title]{Linearly Homomorphic Signature with Tight Security on Lattice}

	\author[1,2]{\fnm{Heng} \sur{Guo}}\email{guoheng@ruc.edu.cn}
	\equalcont{These authors contributed equally to this work.}

	\author[3]{\fnm{Fengxia} \sur{Liu}}\email{shunliliu@gbu.edu.cn}
	
	\author*[1,3]{\fnm{Kun} \sur{Tian}}\email{tkun19891208@ruc.edu.cn}
	\equalcont{These authors contributed equally to this work.}
	
	\author[1,3,4]{\fnm{Zhiyong} \sur{Zheng}}\email{zhengzy@ruc.edu.cn}
	
	\affil[1]{\orgdiv{School of Mathematics}, \orgname{Renmin University of China}, \orgaddress{ \city{Beijing},  \country{China}}}
	
	\affil[2]{\orgdiv{Institute of Interdisciplinary Science}, \orgname{Renmin University of China}, \orgaddress{ \city{Beijing},  \country{China}}}
	
	\affil[3]{\orgname{Great Bay University}, \orgaddress{ \city{Dongguan}, \state{Guangdong Province}, \country{China}}}

	\affil[4]{\orgdiv{Institute of Mathematics}, \orgname{Henan Academy of Sciences}, \orgaddress{ \city{Zhengzhou}, \state{Henan Province}  \country{China}}}

	\footnotetext{\quad $^{1}$This work was supported by Information Security School-Enterprise Joint Laboratory (Dongguan Institute for Advanced Study, Great Bay University, NO.H24120002)and Major Project of Henan Province (No.225200810036).}
	
	
	\abstract{Constructing cryptographic schemes with tight or almost-tight security has long been one of the central problems in theoretical cryptography. At ASIACRYPT 2016, Boyen and Li posed an open problem: whether it is possible to construct a homomorphic signature scheme with tight or almost-tight security under the Short Integer Solution (SIS) assumption in the standard model. In 2024, Chen achieved the first construction with almost-tight security under a weaker security model. To further achieve tight security in the standard model, this paper introduces a new security model whose security requirements are weaker than those of the standard adaptive model but stronger than the model adopted by Chen. Under this model, we construct a linearly homomorphic signature scheme with tight security.}

	
	
	
	\maketitle

	\section{Introduction}\label{sec1}
	
	Homomorphic signatures refer to a type of signature that allows any entity to perform homomorphic operations on authenticated data, enabling the generation of new data. The entity can then obtain a valid signature for the new data without using the signing private key. Since homomorphic signatures were put forward, they have attracted more and more attention. The unique properties of homomorphic signatures provide a vast theoretical research space and significant scientific research value, playing an important role in many practical applications. For example, they can offer effective solutions in fields such as network coding\cite{1,2,3,34,35} and cloud computing\cite{4,5,36,37,38}.
	
	The concept of homomorphic signature schemes was first introduced by R. Rivest \cite{6} in a presentation at Cambridge University in 2000. Shortly thereafter, in 2002, Johnson et al. \cite{7} provided the formal definition and security analysis of homomorphic signatures. Since then, various types of homomorphic signature schemes have emerged. When classifying homomorphic signature schemes based on the homomorphic operation functions, they can be categorized into linear homomorphic signatures, polynomial function homomorphic signatures, and fully homomorphic signatures.

    In 2007, Zhao et al. proposed the first linearly homomorphic signature scheme \cite{1}, which allows arbitrary linear combinations of signed data, facilitating the verification of message integrity and effectively preventing contamination attacks on applications based on network coding. In 2009, Boneh et al. \cite{8} provided the formal definition and security model for linearly homomorphic signature schemes and constructed a linearly homomorphic signature scheme for packet authentication in network coding protocols based on bilinear groups. Subsequently, further improvements were made in terms of efficiency, security, and privacy protection, leading to the proposal of numerous efficient and practical linearly homomorphic signature schemes. Early research primarily based linearly homomorphic signatures on number-theoretic assumptions, such as the discrete logarithm problem or the RSA problem \cite{8,9,10,11,12,39,40,41}. These schemes are efficient but must be designed over $\mathbb{F}_{p}$ (where $p$ is a large prime) to ensure the difficulty of the RSA or discrete logarithm problem, and they cannot resist quantum computing attacks.

To resist quantum attacks, Boneh and Freeman\cite{13} first proposed a linearly homomorphic signature scheme over $\mathbb{F}_{2}$ (the binary field) at the PKC 2011 conference, with its security based on the $k$-SIS problem ($k$-small integer solution problem). Subsequently, at the EUROCRYPT 2011 conference, Boneh and Freeman\cite{14} introduced another linearly homomorphic signature scheme, whose security relies on the hardness of the SIS problem, and proposed a homomorphic signature scheme for polynomial functions, with security based on ideal lattices. In 2013, Wang et al.\cite{15} improved the first linearly homomorphic signature scheme based on lattices, resulting in a new linearly homomorphic signature whose security is directly based on the standard SIS problem. All these schemes\cite{13,14,15} have been proven secure in the random oracle model.
	
	However, in reality, the ideal random oracle model is difficult to achieve. Chen et al. combined bonsai tree techniques\cite{16} with the intersection method of two integer lattices proposed in  [14], and for the first time constructed a lattice-based linearly homomorphic signature scheme over small fields in the standard model\cite{17}. Building on Chen's work, Lin et al. constructed two lattice-based linearly homomorphic signature schemes in the standard model by utilizing full-rank differential hash functions and linearly homomorphic chameleon hash functions\cite{18}.

In the security reduction of a cryptographic system, if the loss in the adversary's advantage is only a constant factor, the system is said to be tightly secure; if the security loss is a logarithmic factor of the security parameter, it is referred to as almost tightly secure(as discussed in Section 4). Tight security is a highly desirable property in the design of cryptographic schemes, as it provides stronger practical security guarantees. Regarding this security property, Boyen and Li \cite{19} posed an open problem at the 2016 Asiacrypt conference (ASIACRYPT 2016): How to construct a short, homomorphic signature scheme with tight or almost tight security in the standard model, based on the hardness of the Short Integer Solution (SIS) problem. In 2024, Chen et al. \cite{20} designed the first lattice-based almost tightly secure linearly homomorphic signature scheme by building upon the scheme in literature \cite{17}, incorporating pseudorandom function techniques\cite{19} and the key-homomorphic  algorithm Eval  \cite{21,22}.  It should be noted, however, that the scheme by Chen et al. achieves almost-tight security only under a security model weaker than the standard adaptive one. How to construct schemes with tight or almost-tight security under stronger security models, or even the standard adaptive security model, remains an important research direction requiring continuous exploration in the field of cryptographic theory.

	\subsection{Our Contribution}
Based on the aforementioned research status, this chapter aims to address the aforementioned open theoretical problem. Specifically, we first construct a security model that is stronger than the one defined by Chen. Under this new model, we further propose a novel linearly homomorphic signature scheme and rigorously prove in the standard model that the scheme achieves tight security under the aforementioned security model, while also satisfying weak context hiding.

\section{ Preliminaries} 
    
\subsection{Notation}

We use the notations $\mathcal{O}$, $\widetilde{\mathcal{O}}$, and $\omega$ to characterize the growth of functions. Let $f$ and $g$ be two functions of $n$. We say that $f(n) = \mathcal{O}(g(n))$ if and only if there exist a constant $c$ and an integer $N$ such that $f(n) \leq c g(n)$ for all integers $n > N$. Similarly, $f(n) = \widetilde{\mathcal{O}}(g(n))$ holds if and only if $f(n) = \mathcal{O}(g(n) \cdot \log^{c'} n)$ for some constant $c'$.
	$f(n)=\omega(g(n))$ if and only if there exists an integer $N$ such that $g(n)\leq cf(n)$ for any constant $c>0$ and any integer $n>N$. Let $n$ denote the security parameter. If $f(n)=\mathcal{O}(n^{c})$ for a constant $c>0$, then we denote $f(n)=\mathrm{poly}(n)$.
	If for all $c>0$, the function $f(n)$ is $\mathcal{O}(n^{-c})$, then this function is called negligible and denoted as $\mathrm{negl}(n)$. If the probability of an event occurring is $1-\mathrm{negl}(n)$, we say that the event holds with overwhelming probability.

	The capital letter $\mathbb{Z}$ denotes the ring of integers, $\mathbb{Z}_{q}$ denotes the ring of integers modulo
	$q(q\geq2)$ and $\mathbb{R}^{n}$ denotes the $n$-dimensional Euclidean space. We then let capital letters $A, B, C$, etc. represent matrices, and bold lowercase letters $\mathbf{a}, \mathbf{b}, \mathbf{c}$ etc. represent vectors. If $A=(\mathbf{a}_{1},...,\mathbf{a}_{n})\in \mathbb{R}^{h\times n}$, the norm of the matrix $A$ is defined as $\parallel A \parallel = \max_{1 \leq i \leq n} \parallel \mathbf{a}_{i} \parallel$, where $\parallel \mathbf{a}_{i} \parallel$ represents the $l_{2}$ norm of the vector. We further denote $\widetilde{A}=(\widetilde{\mathbf{a}}_{1},...,\widetilde{\mathbf{a}}_{n})$ as the Gram-Schmidt orthogonalization of $A$, that is,
	$$\widetilde{\mathbf{a}}_{1}= \mathbf{a}_{1} , \quad \widetilde{\mathbf{a}}_{i}= \mathbf{a}_{i}-\sum_{j=1}^{i-1}\frac{\langle\mathbf{a}_{i}, \widetilde{\mathbf{a}}_{j}\rangle}{\langle \widetilde{\mathbf{a}}_{j}, \widetilde{\mathbf{a}}_{j}\rangle} \widetilde{\mathbf{a}}_{j},  2\leq i\leq n,$$
where $\langle,\rangle$ represents the standard inner product in Euclidean space.
	
	For any distribution $\mathcal{D}$,  $x\sim \mathcal{D}$ means that $x$ follows the distribution $\mathcal{D}$, and $x\leftarrow \mathcal{D}$ means that the sampling of a random value by the distribution. For a set $\mathcal{ X}$, let $x \stackrel{\$}{\leftarrow} \mathcal{ X}$ denote uniformly randomly choosing $x$ form $\mathcal{ X}$. For any probabilistic polynomial-time (PPT) algorithm $Alg $, we write $y\leftarrow Alg(x)$ to denote that the algorithm takes input $x$ and then outputs the result $y$.

\subsection{Linearly homomorphic signature: definition and security model}

Unlike the traditional definition of digital signatures, a linearly homomorphic signature scheme consists of five probabilistic polynomial-time algorithms 
\((\textsf{HSetup}, \textsf{HKeyGen}, \textsf{HSign}, \textsf{Combine}, \textsf{HVerify})\). 
It allows the signer to sign multiple distinct datasets (each associated with a tag), and the size of each dataset may vary. 
It is also assumed that the verifier knows the tag corresponding to the dataset whose computation they wish to verify. 
The definition of a linearly homomorphic signature scheme is as follows:
	
	\begin{definition}[Linearly Homomorphic Signature Scheme]\label{d2.1}
	\normalfont
	A linearly homomorphic signature scheme is a tuple of probabilistic polynomial-time algorithms 
	\((\textsf{HSetup}, \textsf{HKeyGen}, \textsf{HSign}, \textsf{Combine}, \textsf{HVerify})\) defined as follows:
	
	\begin{enumerate}
		\item[$\bullet$] \textsf{HSetup}\((1^{\lambda}, k_0)\): This algorithm takes as input a security parameter \(\lambda\) and a maximum dataset size \(k_0\), and outputs the public parameters \(\textsf{pp}\). The public parameters \(\textsf{pp}\) define the message space \(\mathcal{M}\) and the signature space \(\Sigma\).
		
		\item[$\bullet$] \textsf{HKeyGen}\((\textsf{pp})\): Taking the public parameters \(\textsf{pp}\) as input, this algorithm outputs a public/secret key pair \((\textsf{pk}, \textsf{sk})\).
		
		\item[$\bullet$] \textsf{HSign}\((\textsf{sk}, \tau, \boldsymbol{m})\): Given the secret key \(\textsf{sk}\), a tag \(\tau \in \{0,1\}^h\), and a message \(\boldsymbol{m} \in \mathcal{M}\), this algorithm outputs a signature \(\sigma \in \Sigma\).
		
		\item[$\bullet$] \textsf{Combine}\((\tau, \{(c_i, \sigma_i)\}_{i=1}^{\ell})\): Given a tag \(\tau\) and a set of tuples \(\{(c_i, \sigma_i)\}_{i=1}^{\ell}\), where \(c_i \in \mathbb{F}_q\), \(1 \leq \ell \leq k_0\), and \(\sigma_i \leftarrow \textsf{HSign}(\textsf{sk}, \tau, \boldsymbol{m}_i)\), this algorithm outputs a signature \(\sigma\) for the message \(\sum_{i=1}^{\ell} c_i \boldsymbol{m}_i\).
		
		\item[$\bullet$] \textsf{HVerify}\((\textsf{pk}, \tau, \boldsymbol{m}, \sigma)\): Given the public key \(\textsf{pk}\), a tag \(\tau \in \{0,1\}^h\), a message \(\boldsymbol{m} \in \mathcal{M}\), and a signature \(\sigma\), this algorithm outputs either 0 (reject) or 1 (accept).
	\end{enumerate}
	
	For correctness, it is required that for every \((\textsf{pk}, \textsf{sk})\) output by \(\textsf{HSetup}(1^{\lambda}, k_0)\), the following conditions hold:
	\begin{enumerate}
		\item [(1)] For all tags \(\tau \in \{0,1\}^h\) and all messages \(\boldsymbol{m} \in \mathcal{M}\), if \(\sigma \leftarrow \textsf{HSign}(\textsf{sk}, \tau, \boldsymbol{m})\), then 
		\(\mathsf{HVerify}(\textsf{pk}, \tau, \boldsymbol{m}, \sigma) = 1\).
		
		\item [(2)] For all tags \(\tau \in \{0,1\}^h\) and all sets of tuples \(\{(c_i, \sigma_i)\}_{i=1}^{\ell}\), if \(\sigma_i \leftarrow \mathsf{HSign}(\textsf{sk}, \tau, \boldsymbol{m}_i)\) for each \(i\), then 
		\(\textsf{HVerify}(\textsf{pk}, \tau, \sum_{i=1}^{\ell} c_i \boldsymbol{m}_i, \textsf{Combine}(\tau, \{(c_i, \sigma_i)\}_{i=1}^{\ell}))) = 1\).
	\end{enumerate}
\end{definition}
	
	\textbf{Remark 1}: Typically, we require the message dataset $\boldsymbol{m}_{1}, ..., \boldsymbol{m}_{k}$ to be a set of linearly independent vectors.\\
	
In \cite{20}, Chen aims to construct a linearly homomorphic signature scheme with almost-tight security. However, in practice, the scheme fails to achieve almost-tight security under the standard EUF-CMA (i.e., existential unforgeability under adaptive chosen-message attacks). Therefore, inspired by \cite{27,28}, they proposed unforgeability under selective-tag static chosen-message attacks (U-ST-SCMA) and achieved almost-tight security under this security model. This model requires the adversary to submit the target tag and all dataset messages to be signed to the challenger before obtaining the public key. The specific definition of the U-ST-SCMA security model is as follows:

\begin{definition}[U-ST-SCMA,\cite{20}]\label{d2.4}
	\normalfont
	A linearly homomorphic signature scheme $\mathcal{LS}=(\textsf{HSetup}, \textsf{HKeyGen}, \textsf{HSign}, \textsf{Combine}, \textsf{HVerify})$ is said to satisfy unforgeability under selective-tag static chosen-message attacks (U-ST-SCMA) if for all $k_0$, the advantage of any probabilistic polynomial-time adversary $\mathcal{A}$ in the following game is negligible in the security parameter $\lambda$:
	
	\textbf{Initialization (Tag and Data Selection Phase):} Let $q_s$ be the number of datasets. The challenger uniformly samples $(\tau_i \stackrel{\$}{\leftarrow} \{0,1\}^h)_{i\in \{1,\ldots, q_s\}}$ and sends them to the adversary $\mathcal{A}$. Subsequently, $\mathcal{A}$ specifies a series of datasets $(\overrightarrow{\boldsymbol{m}}_i=(\boldsymbol{m}_{i1},\ldots,\boldsymbol{m}_{ik_0})\in \mathcal{M}^{k_0})_{i\in\{1,\ldots, q_s\}}$ and a target tag $\tau^*\in\{\tau_1,\ldots,\tau_{q_s}\}$, and returns them.
	
	$\mathsf{HSetup}$: The challenger runs $\mathsf{HSetup}(1^{\lambda},k_0)$ to obtain $(\textsf{pk},\textsf{sk})$, and gives $\textsf{pk}$ to $\mathcal{A}$.
	
	$\mathsf{Queries}$: The challenger computes $(\sigma_{ij})_{i\in\{1,\ldots, q_s\}, j\in\{1,\ldots, k_0\}}$, and gives the signatures $(\sigma_{ij})_{i\in\{1,\ldots, q_s\}, j\in\{1,\ldots, k_0\}}$ to $\mathcal{A}$.
	
	$\mathsf{Output}$: $\mathcal{A}$ outputs a tag $\tau^{*}\in\{0,1\}^h$, a message $\boldsymbol{m}^{*}\in \mathcal{M}$, a linear function $f\in \mathcal{F}$, and a signature $\sigma^{*}\in\Sigma$. The adversary wins if $\mathsf{HVerify}(\textsf{pk}, \tau^{*},\boldsymbol{m}^{*},\sigma^{*}, f)=1$ and one of the following conditions holds:
	\begin{enumerate}
		\item[(1)] For all $i$, $\tau^{*}\neq \tau_i$ (Type I forgery);
		\item[(2)] There exists some $i\in [q_s]$ such that $\tau^{*}= \tau_i$ but $\boldsymbol{m}^{*}\neq f(\overrightarrow{\boldsymbol{m}}_i)$ (Type II forgery).
	\end{enumerate}
\end{definition}

The advantage of the adversary $\mathcal{A}$ is defined as the probability that $\mathcal{A}$ wins the security game.	
	
If the adversary is only required to specify the target tag (which does not necessarily belong to the set of tags to be queried) before obtaining the public key, and is not required to submit all dataset messages in advance, this allows the adversary to adaptively choose messages to be signed based on the public key information. This model is stronger than the original U-ST-SCMA, as it grants the adversary greater attacking capability. Schemes satisfying this security model achieve unforgeability under selective-tag adaptive chosen-message attacks, abbreviated herein as U-ST-ACMA. The formal definition is as follows:

	\begin{definition}[U-ST-ACMA]\label{d2.5}
	\normalfont
	A linearly homomorphic signature scheme $\mathcal{LS}=(\textsf{HSetup}, \textsf{HKeyGen}, \textsf{HSign}, \textsf{Combine}, \textsf{HVerify})$ is said to satisfy unforgeability under selective-tag adaptive chosen-message attacks (U-ST-ACMA) if for all $k_0$, the advantage of any probabilistic polynomial-time adversary $\mathcal{A}$ in the following game is negligible in the security parameter $\lambda$:
	
	\textbf{Initialization (Tag Selection Phase):} $\mathcal{A}$ specifies a target tag $\tau^* \in \{0,1\}^h$ and returns it.
	
	$\mathsf{HSetup}$: The challenger runs $\mathsf{HSetup}(1^{\lambda},k_0)$ to obtain $(\textsf{pk},\textsf{sk})$, and gives $\textsf{pk}$ to $\mathcal{A}$.
	
	$\mathsf{Queries}$: Let $q_s$ be the number of datasets. The challenger uniformly samples $(\tau_i \stackrel{\$}{\leftarrow} \{0,1\}^h)_{i\in\{1,\ldots,q_s\}}$ and sends them to the adversary $\mathcal{A}$. 
	$\mathcal{A}$ adaptively selects a series of datasets $(\overrightarrow{\boldsymbol{m}}_i=(\boldsymbol{m}_{i1},\ldots,\boldsymbol{m}_{ik_0})\in \mathcal{M}^{k_0})_{i\in\{1,\ldots,q_s\}}$ and sends them to the challenger. The challenger computes $(\sigma_{ij})_{i\in\{1,\ldots,q_s\}, j\in\{1,\ldots,k_0\}}$, and gives the signatures $(\sigma_{ij})_{i\in\{1,\ldots,q_s\}, j\in\{1,\ldots,k_0\}}$ to $\mathcal{A}$.
	
	$\mathsf{Output}$: $\mathcal{A}$ outputs a tag $\tau^*\in\{0,1\}^h$, a message $\boldsymbol{m}^*\in \mathcal{M}$, a linear function $f\in \mathcal{F}$, and a signature $\sigma^*\in\Sigma$. The adversary wins if $\mathsf{HVerify}(\textsf{pk}, \tau^*,\boldsymbol{m}^*,\sigma^*, f)=1$ and one of the following conditions holds:
	\begin{enumerate}
		\item[(1)] For all $i$, $\tau^*\neq \tau_i$ (Type I forgery);
		\item[(2)] There exists some $i\in [q_s]$ such that $\tau^*= \tau_i$ but $\boldsymbol{m}^*\neq f(\overrightarrow{\boldsymbol{m}}_i)$ (Type II forgery).
	\end{enumerate}
\end{definition}

In homomorphic signature schemes, weak context hiding is also an important consideration. Schemes satisfying this property ensure that new signatures derived from existing ones do not reveal additional information about the original data or the computed functions, thereby providing stronger security guarantees in sensitive scenarios such as medical data analysis and privacy-preserving computations. Based on this, the formal definition of weak context hiding is given below \cite{14,20}.

\begin{definition}[Weak Context Hiding, see \cite{14,20}]\label{d2.54}
\normalfont
A linearly homomorphic signature scheme $\mathcal{LS}=(\textsf{HSetup}, \textsf{HKeyGen}, \textsf{HSign}, \textsf{Combine}, \textsf{HVerify})$ is said to be weakly context-hiding if for all $k_0$, the advantage of any probabilistic polynomial-time adversary $\mathcal{A}$ in the following game is negligible in the security parameter $\lambda$ relative to $\frac{1}{2}$:

$\mathsf{HSetup}$: The challenger $\mathcal{C}$ runs $\mathsf{HSetup}(1^{\lambda},k_0)$ to obtain $(\textsf{pk},\textsf{sk})$, and gives $\textsf{pk}$ and $\textsf{sk}$ to $\mathcal{A}$. The public key defines a message space $\mathcal{M}$ and a signature space $\Sigma$.

$\mathsf{Challenge}$: $\mathcal{A}$ outputs $(\overrightarrow{\boldsymbol{m}}_{0},\overrightarrow{\boldsymbol{m}}_{1},f_{1},\ldots,f_{s})$, where $\overrightarrow{\boldsymbol{m}}_{0},\overrightarrow{\boldsymbol{m}}_{1} \in \mathcal{M}^{k}$, and $\langle f_{i}\rangle=(c_{i1},\ldots,c_{ik_0})$. For all $i\in [s]$, the functions $f_{i}$ satisfy
$$ f_{i}(\overrightarrow{\boldsymbol{m}}_{0}) = f_{i}(\overrightarrow{\boldsymbol{m}}_{1}). $$
In response, the challenger $\mathcal{C}$ generates a random bit $b \in \{0,1\}$ and a random tag $\tau \in \{0,1\}^{h}$. It uses the tag $\tau$ to sign the messages in $\overrightarrow{\boldsymbol{m}}_{b}$, obtaining a vector $\overrightarrow{\sigma}$ containing $k_0$ signatures. Then, for $i=1,\ldots,s$, the challenger computes a signature $\sigma_{i} = \mathsf{Combine}(\tau ,\{(c_{ij},\sigma_{ij})\}_{j=1}^{k_0})$ for $f_{i}(\overrightarrow{\boldsymbol{m}}_{b})$. It sends $(\tau, \sigma_{1},\ldots,\sigma_{s})$ to $\mathcal{A}$. Note that the functions $f_{1},\ldots,f_{s}$ can be output adaptively after $\overrightarrow{\boldsymbol{m}}_{0}$ and $\overrightarrow{\boldsymbol{m}}_{1}$ are submitted.

$\mathsf{Output}$: $\mathcal{A}$ outputs a bit $b^{\prime}$. If $b^{\prime}=b$, the adversary $\mathcal{A}$ wins the game. The advantage of $\mathcal{A}$ is the probability that $\mathcal{A}$ wins the game.
\end{definition}

A linearly homomorphic signature scheme is said to be $s$-weakly context-hiding if an adversary cannot win the privacy game after seeing at most $s$ signatures derived from $\overrightarrow{\boldsymbol{m}}_{0}$ and $\overrightarrow{\boldsymbol{m}}_{1}$.

\subsection{Entropy and Statistical Distance}

\begin{definition}[Statistical distance, \cite{29}]\label{d3.45}
Let $M\subset\mathbb{R}^{n}$ be a finite or countable set, and let $X$ and $Y$ be two discrete random variables taking values in $M$. The statistical distance between $X$ and $Y$ is defined as:
$$ \bigtriangleup (X,Y)=\frac{1}{2}\sum_{a\in M}|P\{X=a\}-P\{Y=a\}|.$$
\end{definition}
When the statistical distance between two distributions is less than a negligible value, we say that the two distributions are statistically  close.

\begin{definition}[\cite{25}, \emph{Min-entropy}]\label{d3.44}
For a random variable \( X \), its min-entropy is defined as:
\[
H_{\infty}(X) = -\log\left( \max_{x \in X} \Pr[X = x] \right).
\]

The average conditional min-entropy of a random variable \( X \) conditional on a correlated variable \( Y \) is defined as:
\[
H_{\infty}(X|Y) = -\log\left( \mathbb{E}_{y \in Y} \left\{ \max_{x \in X} \Pr[X = x | Y = y] \right\} \right).
\]
\end{definition}

The optimal probability of an unbounded adversary guessing $X$ given the correlated value $Y$ is $2^{-H_{\infty}(X|Y)}$\cite{25}.

\begin{lemma}[\cite{25}]\label{l3.122}
Let $X$, $Y$ be arbitrarily random variables where the support of $Y$ lies in  $\mathcal{Y}$, Then
$$ H_{\infty}(X|Y)> H_{\infty}(X)-\log(|\mathcal{Y}|). $$
\end{lemma}

\subsection{ Background on Lattices and Hard Problems}	

\begin{definition}[Lattice, \cite{29}]\label{d3.1}
Let $\Lambda\subset \mathbb{R}^{n}$ be a non-empty subset. $\Lambda$ is called a lattice if:

(1) it is an additive subgroup of $\mathbb{R}^{n}$;

(2) there exists a positive constant $\lambda=\lambda(\Lambda)>0$ such that
$$\mathrm{ min}\{\parallel \mathbf{x}\parallel | \mathbf{x}\in\Lambda , \mathbf{x}\neq 0 \}=\lambda.$$  $\lambda$ is called the minimum distance.
\end{definition}

A full-rank $n$-dimensional lattice can also be expressed as a linear combination of a set of basis vectors $B=\{\mathbf{b}_{1},...,\mathbf{b}_{n}\}\subset \mathbb{R}^{n}$:
$$ \Lambda=\mathcal{L}(B)=\{ B\cdot \mathbf{x}=\sum_{i=1}^{n}x_{i}\mathbf{b}_{i}| x=(x_{1},...,x_{n})^{\top}\in \mathbb{Z}^{n}\}.$$
We call $\Lambda^{*}$  the dual lattice of $\Lambda$ if \begin{equation*}
    \Lambda^{*} = \left\{
        \mathbf{y} \in \mathbb{R}^{n}
        \mid 
        \langle \mathbf{y}, \mathbf{x} \rangle \in \mathbb{Z},
        \quad \text{for all } \mathbf{x} \in \Lambda
    \right\}.
\end{equation*}

\begin{definition}[$q$-ary lattices, \cite{14}]\label{d3.2}
Let $A\in \mathbb{Z}_{q}^{n\times m}$, $u\in \mathbb{Z}^{n}$. The following two $q$-ary lattices are defined as:

(1)$\Lambda_{q}^{\bot}=\{ \mathbf{x}\in \mathbb{Z}^{m} | A\cdot \mathbf{x} \equiv 0 (\mathrm{mod} q)   \}$

(2)$\Lambda_{q}^{\mathbf{u}}=\{ \mathbf{y}\in \mathbb{Z}^{m} | A\cdot \mathbf{y} \equiv \mathbf{u} (\mathrm{mod} q)   \}$

The set $\Lambda_{q}^{\mathbf{u}}$ is a coset of  $\Lambda_{q}^{\bot}$ since $\Lambda_{q}^{\mathbf{u}} =\Lambda_{q}^{\bot}+\mathbf{t}$ for any $\mathbf{t}$ such that $A\cdot \mathbf{t}\equiv \mathbf{u} (\mathrm{mod}q)$.
\end{definition}

\begin{definition}[Short integer solution,  \cite{30}]\label{d3.3}
Let $n$, $m$, $q$ be positive integers, with $m=\mathrm{poly(}n)$. Let $A\in \mathbb{Z}_{q}^{n\times m}$ be a uniformly distributed random matrix over $\mathbb{Z}_{q}$, and let $\beta\in \mathbb{R}$ such that $0<\beta<q$. The SIS problem is to find a short integer solution $\mathbf{x}$ satisfying the following condition:
$$ A\cdot \mathbf{x}\equiv 0(\mathrm{mod}q), \quad \mathrm{and}\quad \mathbf{x}\neq0, \parallel \mathbf{x}\parallel\leq\beta.$$
We denote the \textbf{SIS} problem with parameters $\textbf{SIS}_{q,n,m,\beta}$ or $\textbf{SIS}_{q,\beta}$.
\end{definition}

\begin{theorem}[Worst-case to average-case reduction, \cite{31}]\label{t3.1}
In $\mathbb{Z}_{q}^{n\times m}$, $m=poly(n)$, $q\geq\beta\omega(\sqrt{n\log n})$, and $\beta=poly(n)$, if there is a polynomial time algorithm that solves $\textbf{SIS}_{q,\beta}$ with non-negligible probability, then there is a polynomial time algorithm that finds short non-zero vectors, which
are only a $\lambda\geq \widetilde{\mathcal{O}}(\sqrt{n})$ factor longer than the shortest vector, in all lattices of dimension $n$.
\end{theorem}
	
	\subsection{Discrete Gaussian Distributions, Lattice Trapdoor Algorithms, and Sampling Algorithms}	

\begin{definition}[Discrete Gaussian Distribution, \cite{29}]\label{d3.4}
Let $s$ be a positive real number and $\mathbf{c}\in \mathbb{R}^{n}$ be a vector. The Gaussian function centered at $\mathbf{c}$ with parameter $s$ is defined as: $\rho_{s,\mathbf{c}}(\mathbf{x})=e^{-\frac{\pi}{s^{2}}\parallel\mathbf{x}-\mathbf{c}\parallel^{2}}.$ The discrete Gaussian measure $\mathcal{D}_{\Lambda,s, \mathbf{c}}$ defined on the lattice $\Lambda$ is given by:
$$  \mathcal{D}_{\Lambda,s, \mathbf{c}}(\mathbf{x})=\frac{\rho_{s,\mathbf{c}}(\mathbf{x})}{\rho_{s,\mathbf{c}}(\Lambda)},$$
where $\rho_{s,\mathbf{c}}(\Lambda)=\sum_{\mathbf{x}\in \Lambda}\rho_{s,\mathbf{c}}(x)$.
\end{definition}

Micciancio and Goldwasser \cite{32} proved that a full-rank set $S$ in the lattice $\Lambda$ can be transformed into a basis $T$ such that both have similarly low Gram-Schmidt norms.

\begin{lemma}[\cite{32}, Lemma 7.1]\label{l3.1}
Let $\Lambda$ be an $n$-dimensional lattice. There is a deterministic  polynomial-time algorithm that, given an arbitrary basis of and a full-rank set $S=\{s_{1},...,s_{n}\}$ in $\Lambda$, returns a basis $T$ of $\Lambda$ satisfying
$$ \parallel \widetilde{T}\parallel\leq\parallel \widetilde{S}\parallel \text{and} \parallel T\parallel \leq \parallel S\parallel \frac{\sqrt{n}}{2}. $$
\end{lemma}

\begin{theorem}[\cite{24},Theorem 3.1]\label{t3.2}
Let $q\geq3$ be odd, $h$ be a positive integer,  and let $n:=\lceil 6h\log q\rceil.$  There is a probabilistic
polynomial-time algorithm \textbf{TrapGen}$(q,h,n)$ that outputs a pair $(A\in \mathbb{Z}_{q}^{h\times n}, T_{A}\in \mathbb{Z}^{n\times n})$ such that $A$ is statistically close to a uniform rank $h$ matrix in $\mathbb{Z}_{q}^{h\times n}$, and $T_{A}$ is a basis for $\Lambda_{q}^{\bot}(A)$ satisfying
$$ \parallel\widetilde{T_{A}}\parallel\leq \mathcal{O}(\sqrt{h\log q}) \quad\text{and}\quad \parallel T_{A}\parallel \leq \mathcal{O}(h\log q).$$
\end{theorem}

\begin{lemma}[Sampling from Discrete Gaussian Distribution, \cite{31}]\label{l3.3}
Let $q\geq2$ , $A\in \mathbb{Z}_{q}^{n\times m}$ and let $T_{A}$ be basis for $\Lambda_{q}^{\bot}(A)$ and $s\geq \widetilde{T_{A}}\cdot\omega(\sqrt{\log m })$. Then for $\mathbf{c}\in \mathbb{R}^{n}$ and $\mathbf{u}\in \mathbb{Z}^{n}$:
\begin{enumerate}
  \item  There is a  probabilistic polynomial-time algorithm \textbf{SampleGaussian}$(A,T_{A},s,\mathbf{c})$ that outputs $\mathbf{x}\in \Lambda_{q}^{\bot}(A)$ drawn from a distribution statistically close to $\mathcal{D}_{\Lambda_{q}^{\bot}(A),s,\mathbf{c}}$
  \item  There is a  probabilistic polynomial-time algorithm \textbf{SamplePre}$(A,T_{A},\mathbf{u},s)$ that outputs  $\mathbf{x}\in \Lambda_{q}^{\mathbf{u}}(A)$  sampled from a distribution statistically close to  $\mathcal{D}_{\Lambda_{q}^{\mathbf{u}}(A),s}$, whenever $\Lambda_{q}^{\mathbf{u}}(A)$ is not empty.
\end{enumerate}
\end{lemma}

When $s\geq\omega(\sqrt{\log n})$, we denote the Gaussian sampling algorithm over the integer lattice $\mathbb{Z}^{n}$ as $\textbf{SampleDom}(1^{n},s)$.  That is, when $\mathbf{x}\stackrel{\$}{\leftarrow}\textbf{SampleDom}(1^{n},s)$, $\mathbf{x}$ is statistically close to the distribution $\mathcal{D}_{\mathbb{Z}^{n},s}$.  And these preimages have a conditional min-entropy of $\omega(\log n)$\cite{31}.

\begin{lemma}[\cite{31}]\label{l3.4}
Let $n$ and $h$ be positive integers, and $q$  be a prime, such that  $n\geq2h\lg q$. Then for all but a $2q^{-h}$  fraction of all $A\in \mathbb{Z}_{q}^{h\times n}$ and for any $s\geq\omega(\sqrt{\log n})$, the distribution of the $\alpha=A\cdot \mathbf{x }(\mathrm{mod}q)$
is statistically close to uniform over $\mathbb{Z}_{q}^{h}$, where $\mathbf{x}\stackrel{\$}{\leftarrow}\textbf{SampleDom}(1^{n},s)$.
\end{lemma}

\begin{lemma}[\cite{33}]\label{l3.2}(\cite{33})
Let $\Lambda$ be an $n$-dimensional lattice, and $T$ be a basis of the lattice $\Lambda$. If $s\geq \parallel \widetilde{T}\parallel\cdot \omega(\sqrt{\log n})$, then for any $\mathbf{c}\in \mathbb{R}^{n}$, we have:
$$ \mathrm{Pr}\{\parallel \mathbf{x}-\mathbf{c}\parallel>s\sqrt{n}:\mathbf{x}{\leftarrow} \mathcal{D}_{\Lambda,s,\mathbf{c}}\}\leq \mathrm{negl}(n) .$$
\end{lemma}

\begin{definition}[\emph{Smoothing parameter}\cite{33}]\label{d3.5}
For any $n$-dimensional lattice $\Lambda$ and any given $\epsilon>0$, the smoothing parameter of the lattice is defined as
\begin{equation*}
\eta_{\epsilon}(\Lambda)=\min\left\{s > 0 \mid \rho_{\frac{1}{s}}(\Lambda^{*}) < 1 + \epsilon\right\}.
\end{equation*}
\end{definition}

\begin{lemma}[\cite{31}]\label{l3.5}
Let $q\geq3$, $h$ and $n$ be positive integers satisfying $n\geq 2h\lg q$. Then there exists a negligible function $\epsilon(n)$ such that
$\eta_{\epsilon}(\Lambda_{q}^{\bot}(A))<\omega(\sqrt{\log n})$ for all but at most a $q^{-h}$ fraction of $A$ in the $\mathbb{Z}_{q}^{h\times n}$ .
\end{lemma}

\begin{lemma}[\cite{42}]\label{tfsafa}
	For any $n$-dimensional lattice $\Lambda$, $\mathbf{c} \in \text{span}(\Lambda)$, real number $\epsilon \in (0, 1)$, and $s \geq \eta_\epsilon(\Lambda)$, it holds that
	\[
	\underset{\mathbf{x} \sim D_{\Lambda, s, \mathbf{c}}}{\Pr} \left[ \|\mathbf{x} - \mathbf{c}\| > s\sqrt{n} \right] \leq \frac{1 + \epsilon}{1 - \epsilon} \cdot 2^{-n}
	\]

\end{lemma}

\begin{definition}[Norm of a Random Matrix]\label{d3.422}
	\normalfont
	For the $n$-dimensional unit sphere $\mathbb{S}^n = \{ \boldsymbol{x} \in \mathbb{R}^n : \|\boldsymbol{x}\| = 1 \}$, we define $\|\mathbf{R}\| = \sup_{\boldsymbol{x} \in \mathbb{S}^n} \|\mathbf{R}\cdot \boldsymbol{x}\|$, denoted as $s_{\mathbf{R}}$.
\end{definition}  

\begin{lemma}[\cite{45}]\label{l3.44}
	\normalfont
	Let $\mathbf{R}$ be a random matrix sampled uniformly from $\{1, -1\}^{n \times n}$. Then,
	$
	\Pr\!\bigl[\|\mathbf{R}\| > 12\sqrt{n}\bigr] < e^{-n},
	$
	where $\|\mathbf{R}\| = \sup_{\boldsymbol{x} \in \mathbb{S}^n} \|\mathbf{R} \cdot \boldsymbol{x}\|$.
\end{lemma}

\begin{lemma}[\cite{45}]\label{l3.444}
	\normalfont
	Let $\mathbf{R}$ be an $n \times n$ matrix randomly chosen from $\{-1,1\}^{n \times n}$. Then, for all vectors $\boldsymbol{u} \in \mathbb{R}^n$, we have:
	\[
	\Pr\left[\|\mathbf{R} \boldsymbol{u}\| > \|\boldsymbol{u}\| \sqrt{n} \cdot \omega(\sqrt{\log n})\right] < \mathrm{negl}(n).
	\]
\end{lemma}

\begin{lemma}[Left Basis Sampling,  \cite{45}]\label{le.66} 
	\normalfont
	There exists a polynomial-time algorithm \textsf{SampleBasisLeft} that takes as input matrices $\mathbf{A} \in \mathbb{Z}_q^{n \times m_1}$, $\mathbf{M} \in \mathbb{Z}_q^{n \times m_2}$, a short basis $\mathbf{T}_\mathbf{A}$ of the lattice $\Lambda_q^\perp(\mathbf{A})$, and parameters $0$ and $s$; it outputs a short basis $\mathbf{T}_{\mathbf{F}}$ of the lattice $\Lambda_q^\perp(\mathbf{F})$, where $\mathbf{F} = [\mathbf{A} \mid \mathbf{M}]$. The condition $s \geq \|\tilde{\mathbf{T}}_\mathbf{A}\| \omega\Bigl(\sqrt{\log(m_1 + m_2)}\Bigr)$ holds, and $\|\tilde{\mathbf{T}}_{\mathbf{F}}\| = \|\tilde{\mathbf{T}}_\mathbf{A}\|$.
\end{lemma}

\begin{lemma}[Right Basis Sampling,  \cite{45}]\label{le.6} 
	\normalfont
	There exists a polynomial-time algorithm \textsf{SampleBasisRight} that takes as input matrices $\mathbf{A} \in \mathbb{Z}_q^{n \times k}$, $\mathbf{B} \in \mathbb{Z}_q^{n \times m}$, $\mathbb{R} \in \{-1, 1\}^{k \times m}$, a short basis $\mathbf{T}_\mathbf{B}$ of the lattice $\Lambda_q^\perp(\mathbf{B})$, and parameters $0$ and $s$; it outputs a short basis $\mathbf{T}_{\mathbf{F}}$ of the lattice $\Lambda_q^\perp(\mathbf{F})$, where $\mathbf{F} = [\mathbf{A} \mid \mathbf{AR} + \mathbf{B}]$. The condition $s \geq \| \tilde{\mathbf{T}}_\mathbf{B} \| \cdot s_\mathbb{R} \cdot \omega \big( \sqrt{\log m} \big)$ holds, and $\| \tilde{\mathbf{T}}_{\mathbf{F}} \| \leq \| \tilde{\mathbf{T}}_\mathbf{B} \| (s_\mathbb{R} + 1)$.
\end{lemma}

In \cite{45}, Agrawal et al. introduced a special class of hash functions called ``Full-Rank Difference Hash Functions''. This class of functions has the following important property: for any distinct inputs $id \neq id'$, the difference matrix $\Delta = \mathcal{H}(id) - \mathcal{H}(id')$ is a full-rank matrix.

\begin{definition}[Full-Rank Difference Hash,  \cite{45}]
	\normalfont
	Let $n \in \mathbb{Z}$ be a positive integer. A hash function
	\[
	\mathcal{H}:\{0,1\}^n \to \mathbb{Z}_q^{n \times n}
	\]
	(where $q \ge 2$ is a prime) is called a full-rank difference hash function if it satisfies the following two conditions:
	\begin{enumerate}
		\item \textbf{Efficiently Computable}: The function $\mathcal{H}$ can be computed efficiently;
		\item \textbf{Full-Rank Difference}: For any two distinct inputs $id, id' \in \{0,1\}^n$ ($id \neq id'$), the difference matrix
		\[
		\Delta = \mathcal{H}(id) - \mathcal{H}(id')
		\]
		is a full-rank matrix.
	\end{enumerate}
\end{definition}

\subsection{Statistical Properties Related to Discrete Gaussian Distributions}

\begin{lemma}\label{t3.33}(\cite{13})
	$\mathbf{t}_i \in \mathbb{Z}^m$ and $\mathbf{x}_i$ are mutually independent random variables sampled from a Gaussian distribution $D_{\mathbf{t}_i + \Lambda,\sigma}$ over $\mathbf{t}_i + \Lambda$ for $i = 1,2,\cdots,k$ in which $\Lambda$ is a lattice and $\sigma \in \mathbb{R}$ is a parameter. Let $\mathbf{c} = (c_1,\cdots,c_k) \in \mathbb{Z}^k$ and $g = \mathrm{gcd}(c_1,\cdots,c_k)$, $\mathbf{t} = \sum_{i = 1}^{k} c_i \mathbf{t}_i$. If $\sigma > \|\mathbf{c}\| \eta_{\epsilon}(\Lambda)$ for some negligible number $\epsilon$, then $\mathbf{z}= \sum_{i = 1}^{k} c_i \mathbf{x}_i$ statistically closes to $D_{\mathbf{t} + g\Lambda,\|\mathbf{c}\|\sigma}$.
\end{lemma}

\begin{lemma}\label{l3.777}
	Let \(A \in \mathbb{Z}_q^{n\times m}, s > 0, \mathbf{u} \in \mathbb{Z}^{m}\), and \(\Lambda = \Lambda_q^{\mathbf{u}}(A)\). If \(\mathbf{x}\) is sampled from \(\mathcal{D}_{\mathbb{Z}^m, s}\) conditioned on \(A\mathbf{x} \equiv \mathbf{u} \pmod{q}\), then the conditional distribution of \(\mathbf{x}\) is \(\mathcal{D}_{\Lambda,s}\).
\end{lemma}

\begin{proof}
	According to Definition 9, the discrete Gaussian distribution defined on \(\Lambda\) is:
	\[
	\mathcal{D}_{\Lambda, s}(x) = \frac{\rho_{s}(\mathbf{x})}{\rho_{s}(\Lambda)}, \quad \mathbf{x} \in \Lambda, \text{ where } \rho_{s}(\mathbf{x}) = e^{-\frac{\pi}{s^{2}}\|\mathbf{x}\|^{2}}.
	\]
	
	Then, for \(\mathbf{x} \sim \mathcal{D}_{\mathbb{Z}^m, s}\) and \(\mathbf{x} \in \Lambda\), it can be regarded as the following conditional distribution:
	
	\[
	\mathcal{D}_{\mathbb{Z}^m, s}(\mathbf{x} \mid \mathbf{x} \in \Lambda) = \frac{\mathcal{D}_{\mathbb{Z}^m, s}(\mathbf{x})}{\mathcal{D}_{\mathbb{Z}^m, s}(\Lambda)} = \frac{\rho_{s}(x)/\rho_{s}(\mathbb{Z}^m)}{\rho_{s}(\Lambda)/\rho_{s}(\mathbb{Z}^m)} = \frac{\rho_{s}(\mathbf{x})}{\rho_{s}(\Lambda)} = \mathcal{D}_{\Lambda, s}(\mathbf{x}).
	\]
	
	Therefore, \(\mathbf{x} \sim \mathcal{D}_{\Lambda, s}\).
	
\end{proof}

\begin{theorem}\label{gh}
	Assume that $s_{1} \geq s_{2} \geq \eta_{\epsilon}(\mathbb{Z}^{n})$, where $\epsilon = \frac{1}{\mathrm{poly}(n)}$ is a sufficiently small value. If $|\frac{1}{s_{1}^{2}} - \frac{1}{s_{2}^{2}}| \leq \frac{\epsilon}{2\pi n s_{1}^{2}}$ and $(\frac{s_{2}}{s_{1}})^{n} \geq 1 - \frac{\epsilon}{1 - \epsilon}$, then the statistical distance between $D_{\mathbb{Z}^{n},s_{1}}$ and $D_{\mathbb{Z}^{n},s_{2}}$ is at most $3\epsilon$.
\end{theorem}

\begin{proof}
	By the Poisson summation formula, we have:
	
	$$
	\rho_{s}(\mathbb{Z}^{n}) = \sum_{\mathbf{x} \in \mathbb{Z}^{n}} \exp\left(-\pi \frac{\|\mathbf{x}\|^2}{s^2}\right) = s^n \sum_{\mathbf{x} \in \mathbb{Z}^{n}} \exp\left(-\pi \|\mathbf{x}\|^2 s^2\right).
	$$
	
	Since \( s > \eta_{\epsilon}(\mathbb{Z}^{n}) \) and the dual lattice of \( \mathbb{Z}^{n} \) is itself, it follows that:
	$$
	\rho_{s}(\mathbb{Z}^{n}) = s^n \sum_{\mathbf{x} \in \mathbb{Z}^{n}} \exp\left(-\pi \|\mathbf{x}\|^2 s^2\right) = s^n \rho_{1/s}(\mathbb{Z}^{n}) \leq s^n (1 + \epsilon).
	$$ 
	Thus, we obtain the inequality:
	$$
	s_i^n \leq \rho_{s_i}(\mathbb{Z}^{n}) \leq s_i^n (1 + \epsilon).
	$$
	
	Consider the ratio \( \frac{\rho_{s_2}(\mathbb{Z}^{n})}{\rho_{s_1}(\mathbb{Z}^{n})} \). Noting that \( \frac{s_2}{s_1} \leq 1 \), we have:
	$$
	\left( \frac{s_2}{s_1} \right)^n \frac{1}{1 + \epsilon} \leq \frac{\rho_{s_2}(\mathbb{Z}^{n})}{\rho_{s_1}(\mathbb{Z}^{n})} \leq \frac{(1 + \epsilon) s_2^n}{s_1^n} \leq 1 + \epsilon.
	$$
	
	Let \( U = \{ \mathbf{x} \mid \|\mathbf{x}\| \leq 1 \} \) and \( B = \{ \mathbf{x} \mid \|\mathbf{x}\| \leq s_1 \sqrt{n} \} \). For sufficiently small \( \epsilon \) (e.g., \( \epsilon = 1/\text{poly}(n) \)), by Theorem 6 (for large \( n \)), we get:
	$$
	\Pr_{\mathbf{x} \sim \mathcal{D}_{\mathbb{Z}^n, s_1}} \left[ \|\mathbf{x}\| > s_1 \sqrt{n} \right] = \frac{\rho_{s_1}(\mathbb{Z}^n \setminus s_1 \sqrt{n} U)}{\rho_{s_1}(\mathbb{Z}^n)} = \sum_{\mathbf{x} \notin B} D_{\mathbb{Z}^n, s_1}(\mathbf{x}) \leq \frac{1 + \epsilon}{1 - \epsilon} \cdot 2^{-n} \leq \epsilon.
	$$
	
	Similarly, 
	$$
	\sum_{\mathbf{x} \notin B} D_{\mathbb{Z}^n, s_2}(\mathbf{x}) = \frac{\rho_{s_2}(\mathbb{Z}^n \setminus s_1 \sqrt{n} U)}{\rho_{s_2}(\mathbb{Z}^n)} \leq \frac{\rho_{s_2}(\mathbb{Z}^n \setminus s_2 \sqrt{n} U)}{\rho_{s_2}(\mathbb{Z}^n)} \leq \frac{1 + \epsilon}{1 - \epsilon} \cdot 2^{-n} \leq \epsilon.
	$$
	
	Thus, 
	$$
	\sum_{\mathbf{x} \notin B} \left| \mathcal{D}_{\mathbb{Z}^{n}, s_1}(\mathbf{x}) - \mathcal{D}_{\mathbb{Z}^{n}, s_2}(\mathbf{x}) \right| \leq \sum_{\mathbf{x} \notin B} \mathcal{D}_{\mathbb{Z}^{n}, s_1}(\mathbf{x}) + \sum_{\mathbf{x} \notin B} \mathcal{D}_{\mathbb{Z}^{n}, s_2}(\mathbf{x}) \leq 2\epsilon.
	$$
	
	For \( \mathbf{x} \in B \), consider the ratio:
	$$
	\frac{\rho_{s_1}(\mathbf{x})}{\rho_{s_2}(\mathbf{x})} = \exp\left( -\pi \|\mathbf{x}\|^2 \left( \frac{1}{s_1^2} - \frac{1}{s_2^2} \right) \right).
	$$
	
	Notice that:
	$$
	\left| \pi \|\mathbf{x}\|^2 \left( \frac{1}{s_1^2} - \frac{1}{s_2^2} \right) \right| \leq \pi s_1^2 n \cdot \frac{\epsilon}{2\pi n s_1^2} = \frac{\epsilon}{2}.
	$$
	
	Since \( |e^z - 1| \leq 2|z| \) for \( |z| \leq \frac{1}{2} \), we have:
	$$
	1 - \epsilon \leq \frac{\rho_{s_1}(\mathbf{x})}{\rho_{s_2}(\mathbf{x})} \leq 1 + \epsilon.
	$$
	
	For \( \mathbf{x} \in B \), it follows that:
	$$
	\frac{\mathcal{D}_{\mathbb{Z}^{n}, s_1}(\mathbf{x})}{\mathcal{D}_{\mathbb{Z}^{n}, s_2}(\mathbf{x})} = \frac{\rho_{s_1}(\mathbf{x})}{\rho_{s_2}(\mathbf{x})} \cdot \frac{\rho_{s_2}(\mathbb{Z}^{n})}{\rho_{s_1}(\mathbb{Z}^{n})} \subset \left[ \left( \frac{s_2}{s_1} \right)^n \frac{1 - \epsilon}{1 + \epsilon}, (1 + \epsilon)^2 \right] \subset [1 - 3\epsilon, 1 + 4\epsilon].
	$$
	
	Hence, 
	$$
	\left| \mathcal{D}_{\mathbb{Z}^{n}, s_1}(\mathbf{x}) - \mathcal{D}_{\mathbb{Z}^{n}, s_2}(\mathbf{x}) \right| \leq 4\epsilon \mathcal{D}_{\mathbb{Z}^{n}, s_1}(\mathbf{x}),
	$$
	
	and therefore:
	$$
	\sum_{\mathbf{x} \in B} \left| \mathcal{D}_{\mathbb{Z}^{n}, s_1}(\mathbf{x}) - \mathcal{D}_{\mathbb{Z}^{n}, s_2}(\mathbf{x}) \right| \leq 4\epsilon \sum_{\mathbf{x} \in B} \mathcal{D}_{\mathbb{Z}^{n}, s_1}(\mathbf{x}) \leq 4\epsilon.
	$$
	
	Combining these results, we calculate:
	\begin{align*}
		\frac{1}{2} \sum_{\mathbf{x} \in \mathbb{Z}^{n}} &\left| \mathcal{D}_{\mathbb{Z}^{n}, s_1}(\mathbf{x}) - \mathcal{D}_{\mathbb{Z}^{n}, s_2}(\mathbf{x}) \right| \\
		&= \frac{1}{2} \sum_{\mathbf{x} \in B} \left| \mathcal{D}_{\mathbb{Z}^{n}, s_1}(\mathbf{x}) - \mathcal{D}_{\mathbb{Z}^{n}, s_2}(\mathbf{x}) \right| \\
		&\quad + \frac{1}{2} \sum_{\mathbf{x} \notin B} \left| \mathcal{D}_{\mathbb{Z}^{n}, s_1}(\mathbf{x}) - \mathcal{D}_{\mathbb{Z}^{n}, s_2}(\mathbf{x}) \right| \\
		&\leq \frac{1}{2} (4\epsilon + 2\epsilon) \\
		&\leq 3\epsilon.
	\end{align*}
	
\end{proof}

\begin{corollary}\label{co}
	Let $k = \mathrm{poly}(n)$ be even, $q = \mathrm{poly}(n) \geq (nk)^2$, and $V = \sqrt{2n\log q}\log n \geq \omega(\log n)$. Then $\mathcal{D}_{\mathbb{Z}^{n}, V}$ and $\mathcal{D}_{\mathbb{Z}^{n}, \sqrt{\frac{k \pm 2}{k}}V}$ are statistically indistinguishable.
\end{corollary}

\begin{proof}
	Let $s_1 = V$ and $s_2 = \sqrt{\frac{k-2}{k}}V$. We proceed to show that $\mathcal{D}_{\mathbb{Z}^n, s_1}$ and $\mathcal{D}_{\mathbb{Z}^n, s_2}$ are statistically indistinguishable.
	
	By Lemma \ref{l3.5} and the construction of $V$, there exists a sufficiently small $\epsilon$ such that $s_1 > \eta_{\epsilon}(\mathbb{Z}^n)$ and $s_2 > \eta_{\epsilon}(\mathbb{Z}^n)$. For convenience, let $\epsilon = \frac{1}{n^d}$ where $d > 0$ and $k = n^{d+2}$. When $n$ is sufficiently large, we have:
	
	$$
	\left| \frac{1}{s_1^2} - \frac{1}{s_2^2} \right| = \left| \frac{1}{V^2} - \frac{1}{\frac{k-2}{k}V^2} \right| = \frac{2}{(k-2)V^2} \leq \frac{1}{2\pi n^{d+1}V^2} = \frac{\epsilon}{2\pi n s_1^2}.
	$$
	
	Next, applying Bernoulli's inequality: for any $x \in (0,1)$ and $y \geq 1$, $(1-x)^y \geq 1 - xy$. Thus,
	
	\begin{equation}
		\begin{aligned}
			\left( \frac{s_2}{s_1} \right)^n 
			&= \left( 1 - \frac{2}{k} \right)^{n/2} = \left( 1 - \frac{2}{n^{d+2}} \right)^{n/2} \\
			&\geq 1 - \frac{2}{n^{d+2}} \cdot \frac{n}{2} = 1 - \frac{1}{n^{d+1}} \\
			&\geq 1 - \frac{1}{n^d - 1} = 1 - \frac{1}{\frac{1}{\epsilon} - 1} = 1 - \frac{\epsilon}{1 - \epsilon}.
		\end{aligned}
	\end{equation}
	
	By Theorem \ref{gh}, $\mathcal{D}_{\mathbb{Z}^n, V}$ and $\mathcal{D}_{\mathbb{Z}^n, \sqrt{\frac{k-2}{k}}V}$ are statistically indistinguishable.
	
	Following identical steps, we can show that $\mathcal{D}_{\mathbb{Z}^n, V}$ and $\mathcal{D}_{\mathbb{Z}^n, \sqrt{\frac{k+2}{k}}V}$ are also statistically indistinguishable. This completes the proof.
\end{proof}

\section{New linearly homomorphic signature Scheme}

Before introducing the specific construction methods, it is first necessary to propose a decomposition algorithm named Decompose. The detailed content is as follows:
\begin{theorem}[Decomposition Algorithm]\label{de}
	\normalfont
	Let $k$ be an odd integer satisfying $k = \mathrm{poly}(n)$. Define the following sets:
	\begin{align*}
		\mathcal{X} &= \{\boldsymbol{x}\in \mathbb{F}_{2}^{2k} \mid \|\boldsymbol{x}\| = k-1\} \sqcup \{0\}, \\
		\mathcal{Y} &= \{\boldsymbol{y} \in \mathbb{F}_{2}^{2k} \mid \|\boldsymbol{y}\| = k\} \sqcup \{0\}, \\
		\mathcal{Z} &= \{\boldsymbol{z} \in \mathbb{F}_{2}^{2k} \mid \|\boldsymbol{z}\| = k+1\} \sqcup \{0\}.
	\end{align*}
	When $\|\boldsymbol{w}\|\in[2k-1]$, there exists a deterministic polynomial-time algorithm $\mathsf{Decompose}(\boldsymbol{w})$ that takes as input $\boldsymbol{w} \in \mathbb{F}_{2}^{2k}$ and outputs a pair of vectors $(\boldsymbol{u}, \boldsymbol{v})$ satisfying $\boldsymbol{w} = \boldsymbol{u}+\boldsymbol{v}$, where $\boldsymbol{u} \in \mathcal{X}$ and $\boldsymbol{v} \in \mathcal{Y}$ or $\boldsymbol{v} \in \mathcal{Z}$. Furthermore, when $\boldsymbol{w} \neq 0$, at most one of $\boldsymbol{u}$ and $\boldsymbol{v}$ is the zero vector.
\end{theorem}
\begin{proof}
	For any $\mathbf{w} \in \mathbb{F}_{2}^{2k}$, let $\|\mathbf{w}\| = t$. The algorithm proceeds as follows:
	
	\textbf{Case 1: $t$ is odd}
	
	(1) If $t = k$, set $\mathbf{u} = 0$ and $\mathbf{v} = \mathbf{w}$. Clearly, $\mathbf{w} = \mathbf{u} + \mathbf{v}$ with $\mathbf{u} \in \mathcal{X}$ and $\mathbf{v} \in \mathcal{Y}$, so the algorithm outputs $(\mathbf{u}, \mathbf{v})$. 
	
	(2) If $1 \leq t < k$, suppose the coordinates of $\mathbf{w}$ equal to 1 are $i_{1}, i_{2}, \dots, i_{t}$ where $1 \leq i_{s} \leq 2k$ for $s = 1, \dots, t$.
	\begin{itemize}
		\item[$\bullet$] Compute a pair $(n, m)$ such that $t = n + m$ and $n - m = 1$;
		\item[$\bullet$] Construct $\mathbf{v}$: Select $n$ coordinates from $i_{1}, i_{2}, \dots, i_{t}$ and set them to 1. Choose $k - n$ coordinates from the remaining $2k - t$ coordinates and set them to 1. All other coordinates are 0. Denote this vector as $\mathbf{v}$;
		\item[$\bullet$] Construct $\mathbf{u}$: Set the remaining $m$ coordinates in $i_{1}, i_{2}, \dots, i_{t}$ (not selected for $\mathbf{v}$) to 1, and set the $n$ selected coordinates to 0. All other coordinates match $\mathbf{v}$. Denote this vector as $\mathbf{u}$.
	\end{itemize}
	Verification: Clearly, $\mathbf{w} = \mathbf{u} + \mathbf{v}$. Moreover, $\|\mathbf{u}\| = m + k - n = k - (n - m) = k - 1$ and $\|\mathbf{v}\| = n + (k - n) = k$, so $\mathbf{u} \in \mathcal{X}$ and $\mathbf{v} \in \mathcal{Y}$. The algorithm outputs $(\mathbf{u}, \mathbf{v})$.
	
	(3) If $k < t \leq 2k - 1$, suppose the coordinates of $\mathbf{w}$ equal to 1 are $i_{1}, i_{2}, \dots, i_{t}$.
	\begin{itemize}
		\item[$\bullet$] Compute a pair $(n, m)$ such that $t = n + m$ and $n - m = 1$;
		\item[$\bullet$] Construct $\mathbf{v}$: Select $n$ coordinates from $i_{1}, i_{2}, \dots, i_{t}$ and set them to 1. Choose $\frac{2k - 1 - t}{2}$ coordinates from the remaining $2k - t$ coordinates and set them to 1. All other coordinates are 0. Denote this vector as $\mathbf{v}$;
		\item[$\bullet$] Construct $\mathbf{u}$: Set the remaining $m$ coordinates in $i_{1}, i_{2}, \dots, i_{t}$ (not selected for $\mathbf{v}$) to 1, and set the $n$ selected coordinates to 0. All other coordinates match $\mathbf{v}$. Denote this vector as $\mathbf{u}$.
	\end{itemize}
	Verification: Clearly, $\mathbf{w} = \mathbf{u} + \mathbf{v}$. Moreover,
	\begin{align*}
		\|\mathbf{u}\| &= m + \frac{2k - 1 - t}{2} = m + \frac{2k - 1 - (n + m)}{2} = k - 1, \\
		\|\mathbf{v}\| &= n + \frac{2k - 1 - t}{2} = n + \frac{2k - 1 - (n + m)}{2} = k,
	\end{align*}
	so $\mathbf{u} \in \mathcal{X}$ and $\mathbf{v} \in \mathcal{Y}$. The algorithm outputs $(\mathbf{u}, \mathbf{v})$.
	
	\textbf{Case 2: $t$ is even}
	
	(1) If $t = 0$, set $\mathbf{u} = \mathbf{w}$ and $\mathbf{v} = \mathbf{w}$. This trivially satisfies the output conditions, so the algorithm outputs $(\mathbf{u}, \mathbf{v})$.
	
	(2) If $t = 2k$, randomly select a non-zero  vector $\mathbf{u} \in \mathcal{X}$ and set $\mathbf{v} = \mathbf{w} + \mathbf{u}$. Then $\mathbf{w} = \mathbf{u} + \mathbf{v}$ and $\|\mathbf{v}\| = 2k - (k - 1) = k + 1$, so $\mathbf{v} \in \mathcal{Z}$. The algorithm outputs $(\mathbf{u}, \mathbf{v})$.
	
	(3) If $t = k - 1$, set $\mathbf{u} = \mathbf{w}$ and $\mathbf{v} = 0$. This satisfies the output conditions, so the algorithm outputs $(\mathbf{u}, \mathbf{v})$.
	
	(4) If $t = k + 1$, set $\mathbf{u} = 0$ and $\mathbf{v} = \mathbf{w}$. This satisfies the output conditions, so the algorithm outputs $(\mathbf{u}, \mathbf{v})$.
	
	(5) If $2 \leq t < k - 1$, suppose the coordinates of $\mathbf{w}$ equal to 1 are $i_{1}, i_{2}, \dots, i_{t}$.
	\begin{itemize}
		\item[$\bullet$] Compute a pair $(n, m)$ such that $t = n + m$ and $n - m = 2$;
		\item[$\bullet$] Construct $\mathbf{v}$: Select $n$ coordinates from $i_{1}, i_{2}, \dots, i_{t}$ and set them to 1. Choose $k + 1 - n$ coordinates from the remaining $2k - t$ coordinates and set them to 1. All other coordinates are 0. Denote this vector as $\mathbf{v}$;
		\item[$\bullet$] Construct $\mathbf{u}$: Set the remaining $m$ coordinates in $i_{1}, i_{2}, \dots, i_{t}$ (not selected for $\mathbf{v}$) to 1, and set the $n$ selected coordinates to 0. All other coordinates match $\mathbf{v}$. Denote this vector as $\mathbf{u}$.
	\end{itemize}
	Verification: Clearly, $\mathbf{w} = \mathbf{u} + \mathbf{v}$. Moreover, $\|\mathbf{u}\| = m + k + 1 - n = k + 1 - (n - m) = k - 1$ and $\|\mathbf{v}\| = n + k + 1 - n = k + 1$, so $\mathbf{u} \in \mathcal{X}$ and $\mathbf{v} \in \mathcal{Z}$. The algorithm outputs $(\mathbf{u}, \mathbf{v})$.
	
	(6) If $k + 1 < t \leq 2k - 2$, suppose the coordinates of $\mathbf{w}$ equal to 1 are $i_{1}, i_{2}, \dots, i_{t}$.
	\begin{itemize}
		\item[$\bullet$] Compute a pair $(n, m)$ such that $t = n + m$ and $n - m = 2$;
		\item[$\bullet$] Construct $\mathbf{v}$: Select $n$ coordinates from $i_{1}, i_{2}, \dots, i_{t}$ and set them to 1. Choose $\frac{2k - t}{2}$ coordinates from the remaining $2k - t$ coordinates and set them to 1. All other coordinates are 0. Denote this vector as $\mathbf{v}$;
		\item[$\bullet$] Construct $\mathbf{u}$: Set the remaining $m$ coordinates in $i_{1}, i_{2}, \dots, i_{t}$ (not selected for $\mathbf{v}$) to 1, and set the $n$ selected coordinates to 0. All other coordinates match $\mathbf{v}$. Denote this vector as $\mathbf{u}$.
	\end{itemize}
	Verification: Clearly, $\mathbf{w} = \mathbf{u} + \mathbf{v}$. Moreover,
	\begin{align*}
		\|\mathbf{u}\| &= m + \frac{2k - t}{2} = m + \frac{2k - (n + m)}{2} = k - 1, \\
		\|\mathbf{v}\| &= n + \frac{2k - t}{2} = n + \frac{2k - (n + m)}{2} = k + 1,
	\end{align*}
	so $\mathbf{u} \in \mathcal{X}$ and $\mathbf{v} \in \mathcal{Z}$. The algorithm outputs $(\mathbf{u}, \mathbf{v})$.
	
	In all cases, the algorithm finds a valid decomposition $(\mathbf{u}, \mathbf{v})$ in polynomial time.
\end{proof}

\subsection{Construction}

In this paper, the message space is $\mathcal{M}=\mathbb{F}_{2}^{k}$. The sets $\mathcal{X}$, $\mathcal{Y}$, and $\mathcal{Z}$ are defined as in Theorem \ref{de}.

Let $\langle f\rangle=(c_{1},\ldots,c_{k_0})$ denote the linear function $f(\boldsymbol{m}_{1},\ldots,\boldsymbol{m}_{k_0})=c_{1}\boldsymbol{m}_{1}+\ldots+c_{k_0}\boldsymbol{m}_{k_0}$, where $\boldsymbol{m}_{i}\in \mathbb{F}_{2}^{k}$ and $c_{i}\in\mathbb{F}_{2}$ for all $i$.

The linearly homomorphic signature scheme $\mathcal{LS}=(\textsf{HSetup}, \textsf{HKeyGen}, \textsf{HSign}, \textsf{HVerify}, \textsf{Evaluate})$ in this paper is constructed as follows:

$\bullet$\noindent$\mathsf{HSetup}$$(1^\lambda, k_0)$: Takes as input $(\lambda,k_0)$, where $\lambda$ is the security parameter, $k_0=\mathrm{poly}(\lambda)$ is the maximum size of datasets (i.e., the maximum number of linear operations), and $k_0$ is odd. The parameter generation process is described in Algorithm \ref{alg:HSetup}:

\begin{algorithm}[htbp]
	\caption{HSetup}
	\label{alg:HSetup}
	
	\textbf{Input}: Security parameter $\lambda$, maximum dataset size $k_0=\mathrm{poly}(\lambda)$, where $k_0$ is odd
	
	\textbf{Output}: Public parameters \textsf{pp}

\begin{algorithmic}[1]
	\State $k = 2k_0$
	\State $q \geq (nk)^{3}$
	\State $h = \left\lfloor \dfrac{n}{6\log q} \right\rfloor \geq k$
\State $V = \sqrt{2nk\log q} \cdot \log n$
	\State  Select a full-rank difference hash function $\mathcal{H} : \{0,1\}^h \to \mathbb{Z}_q^{h \times h}$
	\Return \textsf{pp}=$(k, q, h, V, \mathcal{H})$
\end{algorithmic}
\end{algorithm}

$\bullet$\noindent\textsf{HKeyGen}(\textsf{pp}): Takes as input the public parameters \textsf{pp}, and outputs a public/secret key pair (\textsf{pk},\textsf{sk}). The complete key extraction process is formally defined in Algorithm \ref{alg: keygen}.  

\begin{algorithm}[htbp]
	\caption{HKeyGen}
	\label{alg: keygen}
	
	\textbf{Input}: Public parameters \textsf{pp}
	
	\textbf{Output}: Public key \textsf{pk}, secret key \textsf{sk}
		\begin{algorithmic}[1]
		\State $(\mathbf{A}, \mathbf{T}_\mathbf{A}) \leftarrow\mathsf{TrapGen}(q, h, n)$. Here $\mathbf{A} \in \mathbb{Z}_q^{h \times n}$ has rank $h$, and $\mathbf{T}_\mathbf{A}$ is a basis of $\Lambda_q^\bot(\mathbf{A})$
		\Repeat
		\State Sample $\bm{\alpha}_1, \dots, \bm{\alpha}_k \leftarrow \mathbb{Z}_q^h$
		\Until {$\bm{\alpha}_1, \dots, \bm{\alpha}_k$ are linearly independent}
		\State  $\mathbf{A}_1, \mathbf{B} \stackrel{\$}{\leftarrow} \mathbb{Z}_q^{h \times n}$
		\State  $\textsf{pk} \gets (\mathbf{A},\mathbf{A}_{1},\mathbf{B},\bm{\alpha}_1, \ldots, \bm{\alpha}_k)$
		\State  $\textsf{sk} \leftarrow  \mathbf{T}_\mathbf{A}$
		\Return $(\textsf{pk}, \textsf{sk})$
	\end{algorithmic}
\end{algorithm}

$\bullet$\noindent\textsf{HSign}$(\textsf{sk}, \tau, \boldsymbol{m})$: Takes as input the secret key $\textsf{sk}=\mathbf{T}_{\mathbf{A}}$, a tag $\tau\in\{0,1\}^{h}$, and a message $\boldsymbol{m}=(m_{1},\ldots,m_{k})\in \mathbb{F}_{2}^{k}$, and outputs a signature $\sigma$. The pseudocode implementation is shown in Algorithm \ref{alg: sign}.  

\begin{algorithm}[htbp]
	\caption{HSign}
	\label{alg: sign}
	
	\textbf{Input}: Secret key $\textsf{sk} = \mathbf{T}_{\mathbf{A}}$, tag $\tau \in \{0,1\}^h$, message $\boldsymbol{m} = (m_{1},\ldots,m_{k}) \in \mathbb{F}_2^k$
	
	\textbf{Output}: Signature $\sigma$
	\begin{algorithmic}[1]
		\State Compute $\mathbf{F}_{\tau}=[\mathbf{A}|\mathbf{A}_{1}+\mathcal{H}(\tau)\mathbf{B}]$
		\State Compute $\mathbf{T}_{\tau} \gets \textsf{SampleBasisLeft}(\mathbf{A}, \mathbf{A}_{1} + \mathcal{H}(\tau)\mathbf{B}, \mathbf{T}_{\mathbf{A}}, 0, V)$
		\State  Compute $(\boldsymbol{u}, \boldsymbol{v}) \gets \mathsf{Decompose}(\boldsymbol{m})$, where $\boldsymbol{m} = \boldsymbol{u} + \boldsymbol{v}$, with $\boldsymbol{u} \in \mathcal{X}$, and $\boldsymbol{v} \in \mathcal{Y}$ or $\mathcal{Z}$
		\If{$\boldsymbol{u} = \mathbf{0}$ or $\boldsymbol{v} = \mathbf{0}$}
	\State Compute $\boldsymbol{t} = \sum_{j=1}^k m_{j} \bm{\alpha}_j$
		\State $\sigma \gets \mathsf{SamplePre}(\mathbf{F}_{\tau}, \mathbf{T}_{\tau}, \boldsymbol{t}, V)$
		\Else
		\State Compute $\boldsymbol{t}(\boldsymbol{u}) = \sum_{j=1}^k u_{j} \bm{\alpha}_j$
		\State  Compute $\boldsymbol{t}(\boldsymbol{v}_i) = \sum_{j=1}^k v_{j} \bm{\alpha}_j$
		\State $\sigma(\boldsymbol{u}) \gets \mathsf{SamplePre}(\mathbf{F}_{\tau}, \mathbf{T}_{\tau}, \boldsymbol{t}(\boldsymbol{u}), V)$
		\State $\sigma(\boldsymbol{v}) \gets \mathsf{SamplePre}(\mathbf{F}_{\tau}, \mathbf{T}_{\tau}, \boldsymbol{t}(\boldsymbol{v}), V)$
		\State $\sigma = \sigma(\boldsymbol{u}) + \sigma(\boldsymbol{v})$
		\EndIf
		\Return  $\sigma$
	\end{algorithmic}
\end{algorithm}

$\bullet$\noindent\textsf{Combine}$(\tau ,\{(c_{i},\sigma_{i})\}_{i=1}^{\ell})$: Takes as input a tag $\tau\in \{0,1\}^{h}$ and a set of tuples $\{(c_{i},\sigma_{i})\}_{i=1}^{\ell}$, where $\sigma_{i}\gets\mathsf{HSign}(\textsf{sk}, \tau, \boldsymbol{m}_{i})$ and $\ell \leq k_{0}$. This algorithm outputs $\sigma=\sum_{i=1}^{k}c_{i}\sigma_{i}$ as the signature for the message $\sum_{i = 1}^{\ell}c_{i}\boldsymbol{m}_{i}$.

$\bullet$\noindent\textsf{HVerify}$(\textsf{pk}, \tau, \boldsymbol{m}, \sigma)$: Takes as input the public key \(\textsf{pk}\), a tag \(\tau \in \{0,1\}^{h}\), a message \(\boldsymbol{m} \in \mathbb{F}_2^{k}\), and a signature \( \sigma \). The verification process is described in Algorithm \ref{alg: verify}. 

\begin{algorithm}[htbp]
	\caption{HVerify}
	\label{alg: verify}
	
	\textbf{Input}: Public key $\textsf{pk} = (\mathbf{A},\mathbf{A}_{1},\mathbf{B} ,\bm{\alpha}_1, \ldots, \bm{\alpha}_k)$, tag $\tau \in \{0,1\}^h$, message $\boldsymbol{m} \in \mathbb{F}_2^k$, signature \( \sigma \)
	
	\textbf{Output}: 1 (valid) or 0 (invalid)
	\begin{algorithmic}[1]
		\State $\mathbf{F}_{\tau}=[\mathbf{A}|\mathbf{A}_{1}+\mathcal{H}(\tau)\mathbf{B}]$
		\State Compute $\boldsymbol{t} = \sum_{j=1}^k m_{i} \bm{\alpha}_j$
		\If{ $\|\sigma\| \leq k \cdot V \cdot \sqrt{kn}$ \textbf{and} $\mathbf{F}_{\tau}\cdot \sigma \pmod{q}\equiv \boldsymbol{t} $}
		\Return $1$ (signature valid)
		\Else
		\Return $0$ (signature invalid)
		\EndIf
	\end{algorithmic}
\end{algorithm}

\begin{remark}
	\normalfont
	In the vector decomposition of a message $\boldsymbol{m}_i$, when one of the components is the zero vector, the proof methodology for the relevant conclusions follows the same approach as in the case where both components are non-zero vectors. Therefore, when proving the correctness and unforgeability of the scheme, it suffices to consider the case where $\boldsymbol{m}_i$ is decomposed into two non-zero vectors.
\end{remark}

\begin{remark}
	\normalfont
	The above parameter settings satisfy the invocation conditions of the \textsf{SampleBasisLeft}, \textsf{SampleBasisRight}, and \textsf{SamplePre} algorithms. We verify this by taking \textsf{SamplePre} and \textsf{SampleBasisRight} as examples:

For the \textsf{SamplePre} algorithm, given $V = \sqrt{2nk\log q} \cdot \log n$ and $\|\widetilde{\mathbf{T}}_{\mathbf{A}}\| \leq \mathcal{O}(\sqrt{h\log q})$ from Theorem \ref{t3.2}, together with $\|\widetilde{\mathbf{T}}_{\tau}\|=\|\widetilde{\mathbf{T}}_{\mathbf{A}}\|$, we have:
\begin{equation*}
	\frac{V}{\|\widetilde{\mathbf{T}}_{\tau}\|} \geq \sqrt{\frac{2nk}{h}} \cdot \log n \geq \sqrt{\log 2n}.
\end{equation*}
That is, $V \geq \|\widetilde{\mathbf{T}}_{\mathrm{ID}}\| \cdot \omega(\sqrt{\log 2n})$, satisfying the condition of Lemma \ref{l3.3}.

For the \textsf{SampleBasisRight} algorithm, by Lemma \ref{l3.44}, when $\mathbf{R} \stackrel{\$}{\leftarrow} \{-1,1\}^{n}$, we have $s_{\mathbf{R}} \leq 12\sqrt{n}$ with overwhelming probability, and:
\begin{equation*}
	\frac{V}{\|\widetilde{\mathbf{T}}_{\mathbf{B}}\| s_{\mathbf{R}}} \geq \frac{1}{12}\sqrt{\frac{2k}{h}} \cdot \log n \geq \sqrt{\log n}.
\end{equation*}
Thus, $V \geq \|\widetilde{\mathbf{T}}_{\mathbf{B}}\| s_{\mathbf{R}} \cdot \omega(\sqrt{\log n})$, satisfying the condition of Lemma \ref{le.6}. The verification for other algorithms is analogous.
\end{remark}

\subsection{Correctness Proof}

We now prove that the proposed scheme in this section satisfies correctness.

\begin{theorem}\label{t4.1}
	\normalfont
	The above linearly homomorphic signature scheme $\mathcal{LS}=(\textsf{HSetup}, \textsf{HKeyGen}, \textsf{HSign}, \textsf{HVerify}, \textsf{Evaluate})$ satisfies correctness with overwhelming probability.
\end{theorem}

\begin{proof}
Following the definition of correctness, we verify it in two cases:

(1) Given $\sigma \gets \textsf{HSign}(\textsf{sk}, \tau, \boldsymbol{m})$, we need to check the following two verification conditions:

By Theorem~\ref{t3.2}, we have $\|\widetilde{\mathbf{T}}_{\mathbf{A}}\| \leq \mathcal{O}(\sqrt{h\log q})$ with overwhelming probability. Since $V = \sqrt{2nk\log q} \cdot \log 2n$, we obtain
\[
\frac{V}{\|\widetilde{\mathbf{T}}_{\mathbf{A}}\|} \geq \sqrt{\frac{2nk}{h}} \cdot \log 2n \geq \sqrt{\log n},
\]
i.e., $V \geq \|\widetilde{\mathbf{T}}_{\mathbf{A}}\| \cdot \omega(\sqrt{\log n})$.

Therefore, by Lemma~\ref{l3.2}, we have $\|\sigma(\boldsymbol{u})\| \leq V \cdot \sqrt{2n}$ and $\|\sigma(\boldsymbol{v})\| \leq V \cdot \sqrt{2n}$ with overwhelming probability. Consequently,
$$\|\sigma\| \leq 2V \cdot \sqrt{2n} < k \cdot V \cdot \sqrt{kn}.$$

Moreover, since
\begin{align*}
	\mathbf{F}_{\tau} \cdot \sigma \pmod{q} 
	&= \mathbf{F}_{\tau} \cdot (\sigma(\boldsymbol{u}) + \sigma(\boldsymbol{v})) \pmod{q} \\
	&= \boldsymbol{t}(\boldsymbol{u}) + \boldsymbol{t}(\boldsymbol{v}) \\
	&= \sum_{j=1}^{k} u_j \bm{\alpha}_j + \sum_{j=1}^{k} v_j \bm{\alpha}_j \\
	&= \sum_{j=1}^{k} (u_j + v_j) \bm{\alpha}_j \\
	&= \sum_{j=1}^{k} m_j \bm{\alpha}_j = \boldsymbol{t},
\end{align*}
the scheme satisfies correctness for a single signature.

(2) Given $(\boldsymbol{m}_1, \ldots, \boldsymbol{m}_\ell) \in (\mathcal{M})^\ell$ and $\overrightarrow{\sigma} = (\sigma_1, \ldots, \sigma_\ell)$, where
$$\sigma_i \gets \textsf{HSign}(\textsf{sk}, \tau, \boldsymbol{m}_i),$$
and $c_i \in \{0,1\}$. By the definition of the \textsf{Combine}($\tau$, $\{(c_i, \sigma_i)\}_{i=1}^{\ell}$) algorithm, $\sigma = \sum_{i=1}^{\ell} c_i \sigma_i$ is the signature for $\sum_{i=1}^{\ell} c_i \boldsymbol{m}_i$. We now verify that $\sigma$ satisfies the two verification conditions:

Since $\|\sigma_i\| \leq 2V \cdot \sqrt{2n}$ holds with overwhelming probability and $|c_i| \leq 1$, we have
$$\|\sigma\| = \|c_1 \sigma_1 + \ldots + c_\ell \sigma_\ell\| \leq \ell \cdot \max_{1 \leq i \leq \ell} \|\sigma_i\| \leq k_0 \cdot V \cdot \sqrt{n} \leq k \cdot V \cdot \sqrt{kn}.$$

Furthermore, since $\sigma_i \gets \mathsf{HSign}(\textsf{sk}, \tau, \boldsymbol{m}_i)$, we have $\mathbf{F}_{\tau} \cdot \sigma_i \pmod{q} = \boldsymbol{t}_i$, where $\boldsymbol{t}_i = \sum_{j=1}^{k} m_{ij} \bm{\alpha}_j$. Therefore,
$$\mathbf{F}_{\tau} \cdot c_i \sigma_i \pmod{q} = c_i \boldsymbol{t}_i.$$
By the linearity of modular operations, we obtain:
\begin{align*}
\mathbf{F}_{\tau} \cdot \sigma \pmod{q}
	&= \mathbf{F}_{\tau} \cdot \left( \sum_{i=1}^{\ell} c_i \sigma_i \right) \pmod{q} \\
	&= \sum_{i=1}^{\ell} \mathbf{F}_{\tau} \cdot c_i \sigma_i \pmod{q} \\
	&= \sum_{i=1}^{\ell} c_i \boldsymbol{t}_i = \sum_{i=1}^{\ell} c_i \left( \sum_{j=1}^{k} m_{ij} \bm{\alpha}_j \right) \\
	&= \sum_{i=1}^{\ell} \left( \sum_{j=1}^{k} c_i m_{ij} \right) \bm{\alpha}_j \\
	&= \sum_{j=1}^{k} \left( \sum_{i=1}^{\ell} c_i m_{ij} \right) \bm{\alpha}_j \\
	&= \sum_{j=1}^{k} m_j \bm{\alpha}_j = \boldsymbol{t}.
\end{align*}

In conclusion, the scheme satisfies correctness with overwhelming probability.
\end{proof}

\subsection{Unforgeability}

\begin{theorem}\label{t5.1}
	\normalfont
	Assume that $k_0 = \mathrm{poly}(\lambda)$, $k = 2k_0$, $h = \left\lfloor \frac{n}{6\log q} \right\rfloor \geq k$, $q \geq (nk)^3$, $V = \sqrt{2nk\log q} \cdot \log n$, and $\beta = 2kV\sqrt{kn} + 24nkV\sqrt{k}$. If the $\mathsf{SIS}_{q,\beta}$ problem is hard, then the above linearly homomorphic signature scheme achieves tight security in the U-ST-ACMA security model.
	
	More specifically, let $\mathcal{A}$ be a probabilistic polynomial-time adversary that makes at most $q_s$ signature queries to the simulator $\mathcal{C}$. If adversary $\mathcal{A}$ breaks the above signature scheme with advantage $\epsilon$ within time $t$, then the simulator $\mathcal{C}$ can construct a probabilistic polynomial-time algorithm $\mathcal{B}$ that outputs a solution to the $\mathsf{SIS}_{q,\beta}$ problem with advantage at least $\epsilon - \mathrm{negl}(n)$ within time $t + \mathcal{O}\big(q_s T_{\mathsf{HSign}} + k T_{\mathsf{SampleDom}}\big)$, where $T_{\mathsf{HSign}}$ denotes the time to generate one simulated signature and $T_{\mathsf{SampleDom}}$ denotes the time for a single execution of the sampling algorithm $\mathsf{SampleDom}$.
\end{theorem}

\begin{proof}
	Suppose there exists a probabilistic polynomial-time adversary $\mathcal{A}$ that breaks the above signature scheme with advantage $\epsilon$ within time $t$. We show that we can construct a simulator $\mathcal{C}$ that uses the adversary $\mathcal{A}$'s forgery capability to build an algorithm $\mathcal{B}$ that outputs a solution to the $\mathsf{SIS}_{q,\beta}$ problem.
	
	Choose a matrix $\mathbf{A} \in \mathbb{Z}_q^{h \times n}$ uniformly at random as an instance of the $\mathsf{SIS}_{q,\beta}$ problem. Note that with overwhelming probability, $\mathbf{A}$ has full row rank.
	
	\begin{enumerate}
		\item \textsf{Initialization}: The adversary $\mathcal{A}$ declares the target tag $\tau^{*}$ and sends it to the challenger $\mathcal{C}$. The challenger $\mathcal{C}$ generates the public key as follows:
		\begin{enumerate}
			\item Run $(\mathbf{B}, \mathbf{T}_{\mathbf{B}}) \leftarrow \textsf{TrapGen}(q, h, n)$, where $\mathbf{B} \in \mathbb{Z}_q^{h \times n}$ and $\mathbf{T}_{\mathbf{B}}$ is the lattice trapdoor.
			\item Choose $\mathbf{R} \stackrel{\$}{\leftarrow} \{-1,1\}^{n \times n}$ uniformly at random, and set $\mathbf{A}_1 = \mathbf{AR} - \mathcal{H}(\tau^{*})\mathbf{B}$ (Note: $\mathbf{A}_1$ is uniformly distributed over $\mathbb{Z}_q^{h \times n}$).
			\item Sample $\boldsymbol{x}_1, \ldots, \boldsymbol{x}_k \leftarrow \textsf{SampleDom}(1^{2n}, V/\sqrt{k/2})$, and set $\bm{\alpha}_i = [\mathbf{A}|\mathbf{AR}] \boldsymbol{x}_i \pmod{q}$ for $i = 1, \ldots, k$.
			\item Set $\textsf{pk} = (\mathbf{A}, \mathbf{A}_1, \mathbf{B}, \bm{\alpha}_1, \ldots, \bm{\alpha}_k)$.
		\end{enumerate}
		Finally, the simulator $\mathcal{C}$ outputs the public key $\textsf{pk}$ to the adversary $\mathcal{A}$.
		
		\item \textsf{Query Phase}: After receiving the public key $\textsf{pk}$, the adversary $\mathcal{A}$ selects a series of datasets $\overrightarrow{\boldsymbol{m}_1}, \ldots, \overrightarrow{\boldsymbol{m}_{q_s}}$ as signature queries, where $\overrightarrow{\boldsymbol{m}}_i = (\boldsymbol{m}_{i1}, \ldots, \boldsymbol{m}_{ik_0}) \in (\mathcal{M})^{k_0}$. For each $i$ ($i = 1, \ldots, q_s$), the simulator $\mathcal{C}$ uniformly chooses $\tau_i$ from $\{0,1\}^h$ and gives the tag $\tau_i$ to $\mathcal{A}$. The challenger $\mathcal{C}$ returns signatures according to the following two cases:
		
		\textbf{Case 1}: $\tau_i \neq \tau^{*}$.
		\begin{enumerate}
			\item Compute $\mathbf{F}_{\tau_i} = [\mathbf{A} \mid \mathbf{A}_1 + \mathcal{H}(\tau_i)\mathbf{B}] = [\mathbf{A} \mid \mathbf{AR} + (\mathcal{H}(\tau_i) - \mathcal{H}(\tau^{*}))\mathbf{B}]$.
			\item Run $\mathbf{T}_{\tau_i} \leftarrow \textsf{SampleBasisRight}(\mathbf{A}, \mathbf{F}_{\tau_i}, \mathbf{R}, \mathbf{T}_\mathbf{B}, 0, V)$.
			\item Compute $(\boldsymbol{u}_{ij}, \boldsymbol{v}_{ij}) \leftarrow \mathsf{Decompose}(\boldsymbol{m}_{ij})$, where $\boldsymbol{m}_{ij} = (m_{ij}^{(1)}, \ldots, m_{ij}^{(k)}) = \boldsymbol{u}_{ij} + \boldsymbol{v}_{ij}$, with $\boldsymbol{u}_{ij} = (u_{ij}^{(1)}, \ldots, u_{ij}^{(k)}) \in \mathcal{X}$ and $\boldsymbol{v}_{ij} = (v_{ij}^{(1)}, \ldots, v_{ij}^{(k)}) \in \mathcal{Y}$ or $\mathcal{Z}$.
			\item Compute $\boldsymbol{t}(\boldsymbol{u}_{ij}) = \sum_{\ell=1}^{k} u_{ij}^{(\ell)} \bm{\alpha}_\ell$ and $\boldsymbol{t}(\boldsymbol{v}_{ij}) = \sum_{\ell=1}^{k} v_{ij}^{(\ell)} \bm{\alpha}_\ell$.
			\item Compute
			$$\sigma(\boldsymbol{u}_{ij}) \gets \mathsf{SamplePre}(\mathbf{F}_{\tau_i}, \mathbf{T}_{\tau_i}, \boldsymbol{t}(\boldsymbol{u}_{ij}), V)$$
			and
			$$\sigma(\boldsymbol{v}_{ij}) \gets \mathsf{SamplePre}(\mathbf{F}_{\tau_i}, \mathbf{T}_{\tau_i}, \boldsymbol{t}(\boldsymbol{v}_{ij}), V).$$
			\item Output the signature $\sigma_{ij} = \sigma(\boldsymbol{u}_{ij}) + \sigma(\boldsymbol{v}_{ij})$, for $1 \leq i \leq q_s$ and $1 \leq j \leq k_0$.
		\end{enumerate}
		
		\textbf{Case 2}: $\tau_i = \tau^{*}$.
		\begin{enumerate}
			\item Compute $\mathbf{F}_{\tau_i} = [\mathbf{A} | \mathbf{AR}]$, which has no trapdoor basis.
			\item Compute $(\boldsymbol{u}_{ij}, \boldsymbol{v}_{ij}) \leftarrow \mathsf{Decompose}(\boldsymbol{m}_{ij})$, where $\boldsymbol{m}_{ij} = (m_{ij}^{(1)}, \ldots, m_{ij}^{(k)}) = \boldsymbol{u}_{ij} + \boldsymbol{v}_{ij}$, with $\boldsymbol{u}_{ij} = (u_{ij}^{(1)}, \ldots, u_{ij}^{(k)}) \in \mathcal{X}$ and $\boldsymbol{v}_{ij} = (v_{ij}^{(1)}, \ldots, v_{ij}^{(k)}) \in \mathcal{Y}$ or $\mathcal{Z}$.
			\item Compute $\sigma(\boldsymbol{u}_{ij}) = \sum_{\ell=1}^{k} u_{ij}^{(\ell)} \boldsymbol{x}_\ell$ and $\sigma(\boldsymbol{v}_{ij}) = \sum_{\ell=1}^{k} v_{ij}^{(\ell)} \boldsymbol{x}_\ell$. By Lemma \ref{t3.33} and Corollary \ref{co}, $\sigma(\boldsymbol{u}_{ij})$ and $\sigma(\boldsymbol{v}_{ij})$ are within negligible statistical distance from the distribution $\mathcal{D}_{\mathbb{Z}^{2n},V}$.
			\item Set $\sigma_{ij} = \sigma(\boldsymbol{u}_{ij}) + \sigma(\boldsymbol{v}_{ij})$, for $1 \leq i \leq q_s$ and $1 \leq j \leq k_0$.
		\end{enumerate}
	\end{enumerate}
	
	We now show that the simulator's output is identically distributed (within negligible statistical distance) to that of the real signature scheme.
	
	In the real scheme, the public key matrix $\mathbf{A}$ is generated by $\mathsf{TrapGen}$, and its distribution is statistically close to uniform over $\mathbb{Z}_q^{h \times n}$. In the simulation, $\mathbf{A}$ is chosen uniformly at random from $\mathbb{Z}_q^{h \times n}$. Therefore, the two distributions are statistically indistinguishable. A similar argument shows that the matrices $\mathbf{A}_1$ and $\mathbf{B}$ are also indistinguishable from those in the real scheme.
	
	Second, $\bm{\alpha}_1, \bm{\alpha}_2, \ldots, \bm{\alpha}_k$ are indistinguishable from their distribution in the real scheme. This follows from Lemma \ref{l3.4}, as $\bm{\alpha}_j = [\mathbf{A}|\mathbf{AR}] \cdot \boldsymbol{x}_j \pmod{q}$ ($j = 1, \ldots, k$) is statistically close to uniform.
	
	Finally, we prove that the simulated signatures are statistically indistinguishable from real signatures. Since the signatures in Case 1 are generated identically to real signatures, we only need to show that the simulated signatures in Case 2 are indistinguishable from real ones.
	
	Since $\boldsymbol{x}_j \stackrel{\$}{\leftarrow} \mathsf{SampleDom}(1^{2n}, s)$ with $s = \frac{V}{\sqrt{k/2}} \geq \omega(\sqrt{\log 2n})$, by Lemma \ref{l3.2}, we have $\|\boldsymbol{x}_j\| \leq s \cdot \sqrt{2n} = \frac{V}{\sqrt{k/2}} \cdot \sqrt{2n}$ with overwhelming probability. Therefore,
	$$\|\sigma(\boldsymbol{u}_{ij})\| = \left\| \sum_{\ell=1}^{k} u_{ij}^{(\ell)} \boldsymbol{x}_\ell \right\| \leq 2V \cdot \sqrt{kn} \leq kV \cdot \sqrt{kn}.$$
	
	Similarly, $\|\sigma(\boldsymbol{v}_{ij})\| \leq V \cdot \sqrt{2kn}$. Hence,
	$$\|\sigma_{ij}\| = \|\sigma(\boldsymbol{u}_{ij}) + \sigma(\boldsymbol{v}_{ij})\| \leq 2V \cdot \sqrt{2kn} \leq kV \cdot \sqrt{kn}.$$
	
	Moreover:
	\begin{equation}
		\begin{aligned}
			\mathbf{F}_{\tau_i} \cdot \sigma(\boldsymbol{u}_{ij}) \pmod{q}
			&= [\mathbf{A}|\mathbf{AR}] \cdot \sum_{\ell=1}^{k} u_{ij}^{(\ell)} \boldsymbol{x}_\ell \pmod{q} \\
			&= \sum_{\ell=1}^{k} u_{ij}^{(\ell)} [\mathbf{A}|\mathbf{AR}] \cdot \boldsymbol{x}_\ell \pmod{q} \\
			&= \sum_{\ell=1}^{k} u_{ij}^{(\ell)} \bm{\alpha}_\ell = \boldsymbol{t}(\boldsymbol{u}_{ij}).
		\end{aligned}
	\end{equation}
	
	Since $\sigma(\boldsymbol{u}_{ij})$ is statistically distributed as $\mathcal{D}_{\mathbb{Z}^{n},V}$ and satisfies $\sigma(\boldsymbol{u}_{ij}) \in \mathcal{L}_1 = \Lambda_q^{\boldsymbol{t}(\boldsymbol{u}_{ij})}(\mathbf{F}_{\tau_i})$, by Lemma \ref{l3.777}, $\sigma(\boldsymbol{u}_{ij})$ is statistically distributed as $\mathcal{D}_{\mathcal{L}_1,V}$. Similarly, $\sigma(\boldsymbol{v}_{ij})$ is statistically distributed as $\mathcal{D}_{\mathcal{L}_2,V}$, where $\mathcal{L}_2 = \Lambda_q^{\boldsymbol{t}(\boldsymbol{v}_{ij})}(\mathbf{F}_{\tau_i})$.
	
	The adversary can also verify:
	\begin{equation*}
		\begin{aligned}
			\mathbf{F}_{\tau_i} \cdot \sigma_{ij} \pmod{q}
			&= \mathbf{F}_{\tau_i} \cdot (\sigma(\boldsymbol{u}_{ij}) + \sigma(\boldsymbol{v}_{ij})) \\
			&= \mathbf{F}_{\tau_i} \cdot \sigma(\boldsymbol{u}_{ij}) + \mathbf{F}_{\tau_i} \cdot \sigma(\boldsymbol{v}_{ij}) \\
			&= \mathbf{F}_{\tau_i} \cdot \sum_{\ell=1}^{k} u_{ij}^{(\ell)} \boldsymbol{x}_\ell + \mathbf{F}_{\tau_i} \cdot \sum_{\ell=1}^{k} v_{ij}^{(\ell)} \boldsymbol{x}_\ell \\
			&= \sum_{\ell=1}^{k} u_{ij}^{(\ell)} [\mathbf{A}|\mathbf{AR}] \cdot \boldsymbol{x}_\ell + \sum_{\ell=1}^{k} v_{ij}^{(\ell)} [\mathbf{A}|\mathbf{AR}] \cdot \boldsymbol{x}_\ell \pmod{q} \\
			&= \sum_{\ell=1}^{k} u_{ij}^{(\ell)} \bm{\alpha}_\ell + \sum_{\ell=1}^{k} v_{ij}^{(\ell)} \bm{\alpha}_\ell \\
			&= \sum_{\ell=1}^{k} (u_{ij}^{(\ell)} + v_{ij}^{(\ell)}) \bm{\alpha}_\ell = \boldsymbol{t}_{ij}.
		\end{aligned}
	\end{equation*}
	
	Therefore, the simulated signatures are statistically indistinguishable from real signatures.
	
	After receiving all the queried signatures, the adversary $\mathcal{A}$ produces a valid forgery $(\tau^{*}, \boldsymbol{m}^{*}, \sigma^{*}, f)$, where $\boldsymbol{m}^{*} = (m_{1}^{*}, \ldots, m_{k}^{*})$ and $\langle f \rangle = (c_{1}, \ldots, c_{k_0})$, such that
	$$\mathsf{Verify}(\textsf{pk}, \tau^{*}, \boldsymbol{m}^{*}, \sigma^{*}, f) = 1.$$
	
	For both Type I and Type II forgeries, the following equality holds:
	$$
	\begin{aligned}
		\mathbf{F}_{\tau^{*}} \cdot \sigma^{*}
		&= [\mathbf{A}|\mathbf{AR}] \cdot \sigma^{*} \pmod{q} \\
		&= t^{*} \\
		&= \sum_{i=1}^{k} m_{i}^{*} \bm{\alpha}_i \\
		&= \sum_{i=1}^{k} m_{i}^{*} [\mathbf{A}|\mathbf{AR}] \cdot \boldsymbol{x}_i \\
		&= [\mathbf{A}|\mathbf{AR}] \cdot \boldsymbol{x}^{*} \pmod{q},
	\end{aligned}
	$$
	where $\boldsymbol{x}^{*} = \sum_{i=1}^{k} m_{i}^{*} \boldsymbol{x}_i$.
	
	By Lemma \ref{l3.2}, with overwhelming probability, $\|\boldsymbol{x}_i\| \leq s \cdot \sqrt{2n} = \frac{V}{\sqrt{k/2}} \cdot \sqrt{2n}$, so $\|\boldsymbol{x}^{*}\| \leq V \cdot \sqrt{2kn} \leq kV \cdot \sqrt{kn}$. From verification condition (1), $\|\sigma^{*}\| \leq k \cdot V \cdot \sqrt{kn}$, hence $\|\sigma^{*} - \boldsymbol{x}^{*}\| \leq 2k \cdot V \cdot \sqrt{kn}$.
	
	Let $\sigma^{*} - \boldsymbol{x}^{*} = [\boldsymbol{e}_1, \boldsymbol{e}_2]^{\top}$. Then:
	$
	\mathbf{A}\boldsymbol{e}_1 + \mathbf{A}\mathbf{R}\boldsymbol{e}_2 = \mathbf{A}(\boldsymbol{e}_1 + \mathbf{R}\boldsymbol{e}_2) = \mathbf{0}.
	$
	Set $\boldsymbol{e} = \boldsymbol{e}_1 + \mathbf{R}\boldsymbol{e}_2$. Its norm satisfies:
	\[
	\|\boldsymbol{e}\| \leq 2kV\sqrt{kn} + 12\sqrt{n} \cdot 2kV\sqrt{kn} = 2kV\sqrt{kn} + 24nkV\sqrt{k} = \beta.
	\]
	
	If $\sigma^{*} - \boldsymbol{x}^{*} \neq \mathbf{0}$, then $\mathcal{C}$ can output $\boldsymbol{e}$ as a non-zero short solution to the $\mathsf{SIS}_{q,\beta}$ problem.
	
	Finally, we discuss the probability that $\sigma^{*} = \boldsymbol{x}^{*}$. By the properties of the sampling algorithm, the preimage has conditional min-entropy $\omega(\log n)$. Therefore,
	$$\Pr[\sigma^{*} = \boldsymbol{x}^{*}] \leq 2^{-\omega(\log n)} \leq \mathrm{negl}(n).$$
	
	In conclusion, there exists an algorithm $\mathcal{B}$ that solves the $\mathsf{SIS}_{q,\beta}$ problem with advantage at least $\epsilon - \mathrm{negl}(n)$ within polynomial time not exceeding $t + \mathcal{O}\big(q_s T_{\mathsf{HSign}} + k T_{\mathsf{SampleDom}}\big)$.
\end{proof}

\textbf{Worst-case to average-case reduction.}  Based on Theorem \ref{t3.1} from the reference \cite{31}, when \(q \geq \beta\cdot\omega(\sqrt{n\log n})\) holds, the computational hardness of the \(\textbf{SIS}_{q,\beta}\) problem is equivalent to that of approximating the SIVP problem in the worst-case setting with an approximation factor of \(\beta\cdot\tilde{O}(\sqrt{n})\). In the Setup algorithm, the requirement \(q\geq(nk)^3\) serves to ensure that \(q\) has a large enough value for Theorem \ref{t3.1} to be applicable.

From the proof of Theorem \ref{t5.1}, it is evident that the above signature scheme also achieves tight security under the U-ST-SCMA security model.

\begin{theorem}\label{t5.2}
	\normalfont
	Assume that $k_0 = \mathrm{poly}(\lambda)$, $k = 2k_0$, $h = \left\lfloor \frac{n}{6\log q} \right\rfloor \geq k$, $q \geq (nk)^3$, $V = \sqrt{2nk\log q} \cdot \log n$, and $\beta = 2kV\sqrt{kn} + 24nkV\sqrt{k}$. If the $\mathsf{SIS}_{q,\beta}$ problem is hard, then the above linearly homomorphic signature scheme achieves tight security under the U-ST-SCMA security model.
\end{theorem}

\subsection{Privacy}

This section proves that the proposed linearly homomorphic signature scheme is weakly context-hiding, as stated in the following theorem.

\begin{theorem}\label{t5.3}
	\normalfont
	Assume that $k_0 = \mathrm{poly}(\lambda)$, $k = 2k_0$, $h = \left\lfloor \frac{n}{6\log q} \right\rfloor \geq k$, $q \geq (nk)^3$, and $V = \sqrt{2nk\log q} \cdot \log n$. Then the above linearly homomorphic signature scheme is weakly context-hiding.
\end{theorem}

\begin{proof}
	The challenger $\mathcal{C}$ runs $\mathsf{HSetup}(1^{\lambda}, k_0)$ to generate a public/secret key pair $(\mathsf{pk}, \mathsf{sk})$ and sends it to the adversary $\mathcal{A}$. Upon receiving the key pair, the adversary $\mathcal{A}$ outputs $(\overrightarrow{\boldsymbol{m}}_{0}, \overrightarrow{\boldsymbol{m}}_{1}, f_{1}, \ldots, f_{s})$ to the challenger $\mathcal{C}$, where $\overrightarrow{\boldsymbol{m}}_{0} = (\boldsymbol{m}_{01}, \ldots, \boldsymbol{m}_{0k_0})$, $\overrightarrow{\boldsymbol{m}}_{1} = (\boldsymbol{m}_{11}, \ldots, \boldsymbol{m}_{1k_0})$, and they satisfy $f_{i}(\overrightarrow{\boldsymbol{m}}_{0}) = f_{i}(\overrightarrow{\boldsymbol{m}}_{1})$ for $i = 1, 2, \ldots, s$.
	
	In response, the challenger $\mathcal{C}$ randomly selects a tag $\tau \in \{0,1\}^h$ and a random bit $b \in \{0,1\}$, signs the messages $\overrightarrow{\boldsymbol{m}}_{b}$ using its secret key to obtain signatures $\sigma_{b1}, \ldots, \sigma_{bk_0}$, and then invokes the algorithm \textsf{Combine}$(\tau, \{(c_{ij}, \sigma_{bj})\}_{j=1}^{k_0})$ to compute
	$$ \sigma_i \leftarrow \textsf{Combine}(\tau, \{(c_{ij}, \sigma_{bj})\}_{j=1}^{k_0}), $$
	where $\langle f_i \rangle = (c_{i1}, \ldots, c_{ik_0})$ for $i = 1, \ldots, s$. Finally, the challenger sends $(\tau, \sigma_{1}, \ldots, \sigma_{s})$ to the adversary $\mathcal{A}$.
	
	We now analyze the probability that the adversary outputs $b' = b$.
	
	According to the signing algorithm, each $\sigma_i$ follows a Gaussian distribution, and its distribution parameters are completely determined by $(\mathbf{F}_{\tau}, V, f_{i}(\overrightarrow{\boldsymbol{m}}_{b}))$. Since $f_{i}(\overrightarrow{\boldsymbol{m}}_{0}) = f_{i}(\overrightarrow{\boldsymbol{m}}_{1})$, the distribution of $\sigma_i$ remains unchanged regardless of the value of $b$. Consequently, no probabilistic polynomial-time adversary can win this privacy game; that is,
	$$ \left|\Pr[b' = b] - \frac{1}{2}\right| = \mathrm{negl}(n). $$
	
	Therefore, the scheme satisfies weak context hiding. This completes the proof.
\end{proof}

\section{Comparison of Related Schemes}

This section presents a comparative analysis between the linearly homomorphic signature scheme proposed in this paper and existing representative lattice-based linearly homomorphic signature schemes. The detailed comparison results are shown in Table \ref{tabs}. The comparison metrics include signature size, public key size, private key size, security model, and tightness of the reduction. Under the premise that all schemes adopt the same sufficiently large security parameter \(n\), let \(r \ll n\) and \(s\) be a polynomial in \(n\).

\begin{table}[htbp]
\centering
\caption{Comparison of lattice-based signature schemes}
\label{tabs}
\begin{tabular}{|c|c|c|c|c|c|}
    \hline
    Scheme & Signature & Public Key & Private Key & Security Model & Tightness \\
    \hline
    \cite{13} & $\mathcal{O}(1)\mathbb{Z}^{2n}$ & $\mathcal{O}(1)\mathbb{Z}_{2q}^{h \times n}$ & $\mathcal{O}(1)\mathbb{Z}^{n \times n}$ & ROM \& EUF-CMA & Non-tight \\
    \hline
    \cite{17} & $\mathcal{O}(1)\mathbb{Z}^{(r+1)n}$ & $\mathcal{O}(r)\mathbb{Z}_{q}^{h \times n} + \mathcal{O}(k)\mathbb{Z}_{q}^{h}$ & $\mathcal{O}(1)\mathbb{Z}_{q}^{n \times n}$ & SM \& EUF-CMA & Non-tight \\
    \hline
    \cite{15} & $\mathcal{O}(1)\mathbb{Z}^{n}$ & $\mathcal{O}(1)\mathbb{Z}_{q}^{h \times n}$ & $\mathcal{O}(1)\mathbb{Z}_{q}^{n \times n}$ & ROM \& EUF-CMA & Non-tight \\
    \hline
    \cite{20} & $\mathcal{O}(1)\mathbb{Z}^{2n}$ & $\mathcal{O}(5+s)\mathbb{Z}_{q}^{h \times n} + \mathcal{O}(k)\mathbb{Z}_{q}^{h}$ & $\mathcal{O}(1)\mathbb{Z}_{q}^{n \times n} + \mathcal{O}(1)\mathbb{Z}_{2}^{s}$ & SM \& U-ST-SCMA & Almost-tight \\
    \hline
    Ours & $\mathcal{O}(1)\mathbb{Z}^{2n}$ & $\mathcal{O}(3)\mathbb{Z}_{q}^{h \times n} + \mathcal{O}(k)\mathbb{Z}_{q}^{h}$ & $\mathcal{O}(1)\mathbb{Z}_{q}^{n \times n}$ & SM \& U-ST-ACMA & Tight \\
    \hline
\end{tabular}
\end{table}

\begin{itemize}
	\item \textbf{Signature Size.} Signature length is a key metric for evaluating the practicality of a scheme. The scheme in \cite{15} has a significant advantage in signature size, with signatures being only a single element in \(\mathbb{Z}^{n}\). However, this efficiency comes at the cost of proving security in the random oracle model, which represents a target that current standard model schemes strive to achieve in terms of signature size. In contrast, the signature lengths in \cite{13,20} and our proposed scheme are \(\mathbb{Z}^{2n}\), while the signature length in \cite{17} is \(\mathbb{Z}^{(r+1)n}\), which is relatively longer. Overall, the signature size of our scheme is comparable to those in \cite{13,20}, outperforms that in \cite{17}, and achieves tight reduction in the standard model.
	
	\item \textbf{Public Key and Private Key Size.} The sizes of the public key and private key reflect the storage overhead of the scheme. Regarding the public key, the schemes in \cite{13,15} have relatively compact public keys (\(\mathcal{O}(1)\) matrices), but their security relies on the random oracle model. The public keys in \cite{17} and \cite{20} are relatively larger, containing \(\mathcal{O}(r)\) or \(\mathcal{O}(s)\) matrices plus \(\mathcal{O}(k)\) vectors. Our scheme achieves a compact public key in the standard model, consisting of only \(\mathcal{O}(3)\) matrices and \(\mathcal{O}(k)\) vectors. Regarding the private key, except for \cite{20} which additionally includes a bit string of size \(\mathcal{O}(s)\), all schemes have private keys of size \(\mathcal{O}(1)\) matrices, and our scheme is comparable in this regard.
	
	\item \textbf{Security Model and Reduction Tightness.} This is the core of our comparative analysis. Early schemes \cite{13,15} only achieve EUF-CMA security in the idealized random oracle model, which has a gap from real-world environments. Although \cite{17} achieves EUF-CMA security in the standard model, its security reduction is not tight. The scheme in \cite{20} is the first to achieve an almost-tight security reduction in the standard model, but its security model is weaker, proving security only in the U-ST-SCMA model, which requires the adversary to pre-specify the target tag and the dataset messages to be signed before obtaining the public key. In contrast, our scheme achieves tight security under the stronger U-ST-ACMA model. Furthermore, as shown in Theorem \ref{t5.2}, our scheme is also tightly secure under the U-ST-SCMA model, demonstrating that our scheme maintains the advantage of tight reduction even under a stronger adversarial attack model.
	
\end{itemize}

In summary, while maintaining provable security in the standard model, our scheme achieves tight security for the first time in the U-ST-ACMA security model. Regarding signature length, our scheme achieves \(\mathbb{Z}^{2n}\), which is larger than that of \cite{15} (\(\mathbb{Z}^{n}\)) in the random oracle model. This represents a common size disadvantage for standard model schemes compared to random oracle model schemes, as standard model constructions require additional algebraic structures or collision-resistant components to eliminate the idealized assumptions of the random oracle, leading to increased signature length. Compared with other standard model schemes, our signature size is comparable to those in \cite{13,20}, but provides stronger and tighter security reductions than \cite{20}. Compared with \cite{17}, our scheme has significant advantages in signature length, public key size, and the strength of the security model.

\section{Conclusion }

Constructing tightly or almost-tightly secure lattice-based homomorphic signature schemes in the standard model has been a long-standing concern in cryptographic theory. Addressing this problem, this paper proposes a novel linearly homomorphic short signature scheme. To achieve a tight security reduction, we define a new security model that is weaker than the standard EUF-CMA model but stronger than the model adopted in \cite{20}. Under this model, our scheme achieves a tight security reduction. Furthermore, we prove that the scheme satisfies weak context hiding. That is, for any two datasets that yield the same output, their corresponding derived signatures are computationally indistinguishable.

Current research still faces several challenges: 
First, whether it is possible to construct a linearly homomorphic short signature scheme in the standard model that satisfies adaptive security (EUF-CMA) with almost-tight or fully tight security; 
The second challenge lies in designing, within the standard model, a linearly homomorphic signature scheme that supports a larger plaintext space (e.g., $\mathbb{F}_q, q>2$) while providing tight or almost-tight security.

More broadly, as noted in \cite{14}, how to design more efficient constructions with tight or almost-tight security for other cryptographic primitives (such as hierarchical encryption, attribute-based encryption, hierarchical signatures, homomorphic signatures supporting polynomial functions, and hierarchical fully homomorphic signatures) remains an open problem in the standard model.


\end{document}